# Effect of grain boundary state and grain size on the microstructure and mechanical properties of alumina obtained by SPS: A case of the amorphous layer on particle surface

M.S. Boldin[1,(*)], A.A. Popov[1], A.V. Nokhrin[1], A.A. Murashov[1], S.V. Shotin[1], V.N. Chuvil'deev[1], N.Yu. Tabachkova[2,3], K.E. Smetanina[1]

[1] Lobachevsky State University of Nizhny Novgorod, Nizhny Novgorod, Russia

[2] National University of Science and Technology "MISIS", Moscow, Russia

[3] A.N. Prokhorov General Physics Institute, Russian Academy of Science, Moscow, Russia

boldin@nifti.unn.ru

**Abstract**

The effect of temperature modes and heating rates ($V_h$) on the shrinkage kinetics of submicron and fine aluminum oxide powders has been studied. The objects of research comprised (i) submicron $\alpha$-$Al_2O_3$ powder, (ii) submicron $\alpha$-$Al_2O_3$ powder with an amorphous layer on particle surface, (iii) fine $\alpha$-$Al_2O_3$ powder. The alumina ceramic specimens were produced by Spark Plasma Sintering (SPS). Equally fine powders (i) and (ii) were used to analyze the effect of an amorphous layer on sintering kinetics. Powders (i) and (iii) were used to analyze the effect of the initial particle size on shrinkage kinetics. Shrinkage curves were analyzed using the Young–Cutler and Coble models. It has been shown that sintering kinetics is determined by the intensity of grain boundary diffusion for submicron powders and by simultaneous lattice and grain boundary diffusion for fine powders. It has been determined that an amorphous layer on the surface of submicron $\alpha$-$Al_2O_3$ powder affects grain boundary migration rate and the Coble equation parameters at SPS final stages. It has been suggested that abnormal characteristics of the alumina ceramics sintered from a submicron powder with an

---

[(*)] corresponding author (boldin@nifti.unn.ru)

amorphous layer on the particle surface are associated with an increased concentration of defects at grain boundaries that were formed during crystallization of the amorphous layer.

## 1. Introduction

Spark plasma sintering (SPS) [1-8] is a process of rapid hot compacting and a powerful method for obtaining ceramic materials with an ultrafine-grained (UFG) microstructure. High heating rates (up to 2500 ºC/min) enable reduction of grain growth intensity, which is important for sintering nano- and submicron powders. Literature analysis demonstrates that SPS can help obtain bulk specimens of high-density ceramics with homogeneous UFG microstructure and, consequently, with excellent mechanical properties [1, 8-13]. This drives an ever-growing interest in the SPS potential and promotes design of new forward-looking industrial systems for SPS [1, 14-15].

The uncertainty about sintering behaviors of nano- and submicron ceramic powders remains one of the key challenges in prospective applications of SPS. Today, researchers agree that the main SPS mechanism in ceramics combines diffusion and creep [1, 7, 16-23]. Some researchers emphasize the role of "spark" and "plasma" processes during SPS [24, 25]. Much attention is also focused on the issue of inhomogeneous electric and thermal fields under fast heating [1, 6, 22, 26, 27]. The authors of [7, 28-31] noted the importance of taking into account the pollution of the metals and ceramics being sintered by carbon, which may improve the mechanical properties of Ti alloys [32], alter the optical characteristics of the oxide ceramics [28-31], change the phase composition of the ceramics and hard alloys based on tungsten carbide [33, 34], etc.

The key challenge for researchers is to be able to achieve quickly the theoretical density in ceramics at low temperatures and short SPS times [1-5, 7, 7, 19, 20]. Up to the present time, the causes of accelerated diffusion resorption of pores in UFG ceramics at low temperatures and short SPS times have not been determined. The studies of the features of rapid SPS of nano- and fine-grained powders are being carried out extensively [35-40]. Much attention is paid to the investigations of the effect of initial particle sizes and of the grain sizes on the densification behavior, on the grain growth, and of the mechanical properties of the ceramics [41, 42].

There are few works on the analysis of the effect of the microstructure of grain boundaries on the mechanical properties of the UFG ceramics. Most works were devoted to the investigations of the effect of small additives of doping elements on the properties of the grain boundaries and, hence, on the properties of the UFG ceramics [43-45]. At the same time, it is worth noting that the rapid heating in SPS allows obtaining the microstructure of the ceramics with the main grain sizes close to the initial ones of the particles [1, 2, 4, 5, 8]. The temperature of obtaining the ceramics by SPS is low while the sintering time is much shorter than in conventional sintering or in hot pressing [1]. It means that the initial particle surface state may also affect essentially the sintering process, the density, the mechanical properties, and the microstructure parameters of the ceramics.

The purpose of this study is to investigate the specifics of rapid compaction of submicron- and fine aluminum oxide powders and to research the effect of microstructural parameters of a alumina ceramic on its mechanical properties. In the presents study, special attention was paid to the analysis of the effect of the grain boundary structure on the sintering kinetics, grain growth, and mechanical properties of the ceramics. The objects of research comprised (i) submicron α-$Al_2O_3$ powder, (ii) submicron α-$Al_2O_3$ powder with an amorphous layer on particle surface, (iii) fine α-$Al_2O_3$ powder. Equally submicron powders (i) and (ii) were used to analyze the effect of an amorphous layer on sintering kinetics. Powders (i) and (iii) were used to analyze the effect of the initial particle size on shrinkage kinetics. The choice of alumina as the object of investigation was motivated by its wide application as a construction material, low cost, and the optimal combination of hardness, fracture toughness, and wear resistance [46-51].

## 2. Materials and techniques

The objects of research comprised three series of industrial α-Al$_2$O$_3$ powders with different degrees of fineness (Table 1).

Table 1. Research object description

| Series | Manufacturer | Initial particle size R$_0$, μm | Composition as per the supplier certificate | Powder specifics revealed in research |
|---|---|---|---|---|
| 1 | Nanoe (France) | ~0.1-0.2 | 100% α-Al$_2$O$_3$ | - |
| 2 | Taimei Chemicals Co., Ltd (Japan) | ~0.2 | 100% α-Al$_2$O$_3$ | Amorphous layer on the particle surface |
| 3 | Alfa Aesar - A Johnson Matthey Company (Germany) | ~1 | 100% α-Al$_2$O$_3$ | Particles with an increased dislocation density |

Dr. Sinter model SPS-625 setup (SPS SYNTEX INC. Ltd., Japan) was used to compact ceramic specimens 12 mm in diameter and 3 mm high through SPS. The applied uniaxial pressure amounted to 70 MPa. The pressure was applied simultaneously with heating. Sintering was done in vacuum (6 Pa). A Chino IR-AH pyrometer focused on the graphite mold surface was used to measure the sintering temperature. Based on previous studies and comparison of the infrared pyrometer readings (T1 in Fig. 1a) and the control thermocouple attached to the alumina specimen surface, T1 values were recalculated into the actual specimen temperature (T2) using the following empirical relation: $T2 = A \cdot T1 - B$, where A and B – experimental constants (Fig. 1a).

Two sintering modes were used in the research:

- mode A: heating at a constant rate ($V_h$ = 10, 50, 100, 250, 350, 700 °C/min) to the shrinkage finishing temperature or to a preset temperature ($T_1$, $T_2$, $T_3$);

- mode B: heating to a preset temperature ($T_1$, $T_2$, $T_3$) at a rate of $V_h$ = 50 °C/min and sintering at this temperature in the isothermal mode (up to 30 min).

Temperature values $T_1$, $T_2$, $T_3$ for powders in Series No. 1-2 were 1320, 1420, 1520 °C and 1470, 1530, 1600 °C for the powder in Series No. 3.

The dependence of effective shrinkage ($L_{eff}$) on sintering times and temperature was observed during SPS. Empty molds were heated during the experiment in order to account for thermal expansion ($L_0$). True shrinkage was calculated as $L = L_{eff} - L_0$ (Fig. 1b). Using L(T) temperature curve in lineralization, shrinkage rate temperature curve was calculated: $S = \Delta L/\Delta t$.

A Sartorius® CPA balance was used to measure the density ($\rho$) of the ceramics by hydrostatic weighing in distilled water at room temperature (RT). The density measurement accuracy was ± 0.005 g/cm³. The theoretical density of $Al_2O_3$ was assumed to be $\rho_{th}$ = 4.05 g/cm³.

A Struers® Duramin-5 hardness tester was used to measure microhardness ($H_v$) at a load of 2 kg. The minimum fracture toughness coefficient $K_{IC}$ was calculated using the Palmquist method based on the length of the maximum radial crack. When calculating $K_{IC}$, the elastic modulus was assumed to be E = 350 GPa. The accuracy of measuring $H_v$ and $K_{1C}$ was ± 1 GPa and ± 0.3 MPa·m$^{1/2}$, respectively. The average values of $H_v$ and $K_{IC}$ were obtained by averaging at least 10 measurements in the central part of sintered ceramic specimens.

X-ray diffraction (XRD) analysis was performed using Shimadzu® XRD-7000 diffractometer (CuK$_\alpha$, the scanning step 0.04°, the exposure 2 s per point). The qualitative phase analysis was carried out using Diffrac.EVA software. The quantitative analysis was carried out by Rietveld method. The initial parameters of the $\alpha$-$Al_2O_3$ phase found in the specimens were taken from the cif-files of ICSD database.

The X-ray Photoemission Spectroscopy (XPS) measurements were performed with PHI Model 5500 ESCA spectrometer using Mg Kα radiation (hν = 1253.6 eV). The vacuum of the analysis chamber was approximately $10^{-8}$ Pa. The powder sample was pressed into an In foil. The sample surface was cleaned with a beam of $Ar^+$ ions with an energy of 2 keV for 30 s at the first stage and then for 6 min. A 2 × 2 $mm^2$ raster provided the sputtering rate of 1–2 nm/min.

A Jeol® JEM 2100 transmission electron microscope (TEM) and Jeol® JSM-6490 and Tescan® Vega 2 scanning electron microscopes (SEM) were used to study the microstructure of specimens. The chord method in conjunction with the GoodGrains software (UNN, Russia) was used to measure the size of particles (R) and grains (d). The average values of R and d were obtained by averaging the results of at least 100 measurements. Studies of the ceramics microstructure (determining d) were performed on polished sections and on specimen fractures. Fracture analysis was performed mainly for small-grain ceramics while section analysis was applied to large-grain ceramics.

For the microstructure investigations, the sintered ceramics were annealed in an air furnace at 750 ºC, 1 h to remove the residual graphite from the surface. Afterwards, the specimen surfaces were subjected to additional mechanical grinding and polishing up to ~50 μm in depth. The polishing allowed removing the surface carbonized layers (see [28-31]). The grinding was performed using Struers® Secotom™ 10 grinding machine, the polishing – using Buehler® Automet™ 250 polishing setup.

### 3. Experimental results

#### 3.1 Powder qualification

Aluminum oxide powder in Series No. 1 has a fairly wide histogram of particle size distribution; as can be seen in Fig. 2a, the powder contains a noticeable number of fairly large particles of 0.3–0.4 μm in size (Fig. 2a). The average particle size $R_1$ was ~ 0.15 μm (Figs. 2b, c). XRD phase analysis did not reveal impurity phases in Series No. 1 powders, with the position of all XRD peaks corresponding to α-Al$_2$O$_3$ phase (Fig. 2d).

Figs. 3a–3c show electron microscopic images of α-Al$_2$O$_3$ powder in Series No. 2 in its initial state. The granulometric composition of the powders is quite homogeneous, large particles were not detected. The particle size of aluminum oxide averages $R_2 \sim 0.2$ μm. An amorphous layer ~5-10 nm thick is detected on the surface of submicron alumina particles (Fig. 3c, d). According to XRD phase analysis, the powder contains no impurity phases (Fig. 2d).

The results of XPS of the surfaces of powders from Series No. 2 are presented in Fig. 4. The XPS concentrations of elements on the surfaces of the powders from Series No. 2 were determined from the overview spectra, which were measured after 30 s sputtering (to remove adsorbed carbohydrates) and after 6 min sputtering. The initial surfaces of the α-Al$_2$O$_3$ particles weren't studied by XPS since the O1s and Al2p (Al2s) peaks wee in different energy bands, and the intensities of these peaks attenuated differently when the adsorbed carbon films are present. After ion sputtering, the estimates can be made because there is no selective sputtering in Al$_2$O$_3$. The O/Al concentrations ration was 1.5-1.52 in both cases. Note that if there were aluminum hydroxide film, this ratio should be greater after 30 s sputtering. The binding energies of Al2p – 74.5 eV and O1s – 531.4 eV are typical for alumina (Fig. 4).

Electron microscopy results reveal a fairly homogeneous granulometric composition of powders in Series No. 3, with the average particle size being $R_3 \sim 0.8$-1 μm (Figs. 5a, b). Most α-Al$_2$O$_3$ particles have an increased dislocation density (Fig. 5c). As with Series No. 1-2, XRD phase analysis suggests that there are no secondary phases in the fine powder of Series No. 3 (Fig. 2d).

## 3.2. Spark Plasma Sintering of powders

### 3.2.1. Sintering Series No. 1 Powder

Figure 6 shows temperature dependences of shrinkage L (Fig. 6a) and shrinkage rate S (Fig. 6b) at various heating rates $V_h$ of submicron aluminum oxide powders in Series No. 1. As can be seen in Fig. 6, the compaction (shrinkage) temperature dependences at all observed heating rates $V_h$ are monotonic up to the point where temperature corresponds to the end of shrinkage. L(T) curves follow a classical three-stage pattern [52]. Increasing the heating rate from 10 to 700 °C/min shifts L(T) shrinkage curves towards higher temperatures by 150–250 °C. The maximum shrinkage $L_{max}$ decreases as the heating rate is growing, which correlates with a decrease in the relative density of the sintered alumina ceramics (Table 2). As can be seen in Table 2, increasing the heating rate $V_h$ from 10 to 700 °C/min causes a monotonic decrease in the relative density ($\rho/\rho_{th}$) of alumina specimens from 99.5% to 98.9%, i.e. by $\Delta\rho/\rho_{th} \sim 0.6\%$.

An analysis of S(T) curves in Fig. 5b demonstrates that increasing the heating rate $V_h$ from 10 to 700 °C brings the maximum shrinkage rate $S_{max} = S(T = T_{max})$ up from $(3-5)\cdot10^{-3}$ to $87\cdot10^{-3}$ mm/s and, simultaneously, the temperature $T_{max}$ from 1260-1320 to ~1460 °C.

Table 2. Microstructural parameters and properties of alumina ceramic specimens produced by SPS from α-$Al_2O_3$ powder of Series No. 1

| Mode | $T_s$, °C | $V_h$, °C/min | t, min | d, μm | $\rho/\rho_{th}$, % | $H_v$, GPa | $K_{IC}$, MPa·m$^{1/2}$ |
|---|---|---|---|---|---|---|---|
| A | 1390 | 10 | 0 | 0.5 | 99.51 | 20.3 | 2.8 |
|   | 1460 | 50 |   | 0.5 | 99.35 | 22.3 | 2.5 |
|   | 1550 | 100 |   | 0.7 | 99.04 | 20.2 | 2.8 |
|   | 1650 | 250 |   | 0.9 | 99.33 | 19.9 | 2.7 |
|   | 1650 | 350 |   | 1.1 | 99.24 | 19.4 | 2.8 |
|   | 1530 | 700 |   | 0.7 | 98.91 | 19.8 | 2.6 |

| | | | 0 | 0.2 | 81.51 | 9.7 | 2.9 |
| --- | --- | --- | --- | --- | --- | --- | --- |
| | | | 1 | 0.2 | 84.23 | 11.3 | 2.7 |
| | | | 2 | 0.2 | 85.44 | 11.8 | 2.6 |
| | | | 3 | 0.2 | 90.51 | 16.1 | 2.6 |
| B | 1350 | 50 | 5 | 0.2 | 94.12 | 18.0 | 2.1 |
| | | | 7 | 0.3 | 94.52 | 18.6 | 2.9 |
| | | | 11 | 0.4 | 96.64 | 20.6 | 2.5 |
| | | | 15 | 0.4 | 97.91 | 21.4 | 2.4 |
| | | | 30 | 0.5 | 99.39 | 21.2 | 2.7 |

The electron microscopy results in Fig. 7 and in Table 2 demonstrate that rapid heating does not lead to a significant change in the average ceramic grain size. No noticeable change in d is apparently due to shorter heating duration (higher heating rate) being compensated by an increase in the temperature to which the specimen is heated. The microstructure of specimens is fairly uniform, and there are no abnormally large grains in the ceramic structure (Fig. 7).

The dependence between the alumina ceramic microhardness and heating rate $H_v(V_h)$ correlates with $d(V_h)$ dependence. Increasing the grain size with the heating rate going from 10 to 350°C/min causes a decrease in microhardness from 20.3 to 19.8 GPa. Note the high microhardness ($H_v$ = 23.2 GPa) of the specimen obtained at a heating rate of $V_h$ = 50 °C/min. Increasing the heating rate has no noticeable effect on fracture toughness of sintered ceramics, with $K_{IC}$ varying from 2.5 up to 2.8 MPa·m$^{1/2}$ (Table 2).

As [52] suggests, intensive grain growth in ceramics starts when the relative density begins to exceed 90%. Research into the compaction kinetics of powders in Mode B was performed in order to evaluate how porosity affects grain growth and microhardness. The powders were heated at a rate of $V_h$ = 50 °C/min to 1350 °C, at which point a relative ceramic density of ~80% was reached, with a subsequent isothermal exposure at 70 MPa. The sintering modes and specimen properties are given in Table 2. Fig. 8 shows the shrinkage curves, Fig. 9 demonstrates the microstructure of ceramic specimens.

As can be seen in Fig. 8a and Table 2, increasing the duration of isothermal exposure from 0 to 30 min causes a monotonic increase in shrinkage to L ~ 1.8 mm and a relative density of ceramics from 81.5% to 99.4%. An increase in the ceramic grain size becomes noticeable only after the specimen relative density reaches 84.5% (t = 7 min). The dependence of ceramic microhardness on isothermal exposure duration correlates $\rho/\rho_{th}(t)$. Sharply increasing the density from 81.5% (t = 0 min) to 96.6% (t = 11 min) leads to an increase in microhardness $H_v$ from 9.6 to 20.6 GPa. Raising the isothermal holding time from 11 to 15 min causes a gradual increase in $\rho/\rho_{th}$ from 96.6% to 97.9% and a minor increase in microhardness from 20.6 to 21.4 GPa. Further expanding the isothermal holding time to 30 min boosts density to 99.4%; with microhardness remaining unchanged. It should be also noted that isothermal exposure has little effect on the minimum fracture toughness coefficient $K_{IC}$ of aluminum oxide.

### 3.2.2. Sintering Series No. 2 Powder

Fig. 10 shows the temperature curves of shrinkage L (Fig. 10a) and shrinkage rate (Fig. 10b) at different heating rates (mode A). These curves suggest that increasing the heating rate from 10 to 700 °C/min shifts L(T) dependences to the region of higher heating temperatures. In particular, if $V_h$ increases from 10 to 700 °C/min, the shrinkage finishing temperature rises from 1265–1280 °C to 1475–1500 °C (Fig. 10a). A similar trend is observed for the dependence between the shrinkage rate and heating temperature where increasing $V_h$ raises the temperature $T_{max}$, at which point the maximum shrinkage rate $S_{max}$ can be observed (Fig. 10b). The maximum shrinkage rate ($L_{max}$) in this case is almost completely independent of the heating rate (Fig. 10a) while $S_{max}$ increases as the heating rate grows (Fig. 10b). Thus, increasing the heating rate intensifies ceramic sintering at the intermediate heating stage (Stage II), yet densification finishes (end of Stage III) at higher temperatures.

Table 3 summarizes the findings of research into density, average grain size, and mechanical properties of ceramics obtained by heating at different rates to temperatures of 1320, 1420, and 1520 °C. These temperatures correspond to the end of the intensive shrinkage stage where the density of sintered ceramics is already high (Table 3) and the pores do not significantly affect the grain boundaries migration mobility.

An analysis of the findings in Table 3 suggests that, at a given heating rate ($V_h$ = const), higher sintering temperatures increase the relative density of alumina. Heating at a rate of 10 °C/min to 1520°C can help produce ceramics with a relative density of 99.72%. Higher heating rates decrease the ceramic density, with the scale of decrease in density ($\Delta\rho$) driven by the heating temperature. As can be seen in Table 3, increasing the heating rate from 10 to 720 °C/min causes the relative density to decrease by 0.44% (from 99.72 to 99.28%) when heated to 1520 °C; by 0.72% (from 99.66 to 98.94%) when heated to 1420°C; and by 7.16% (from 99.46 to 92.30%) when heated to 1320°C. Thus, the heating rate has the biggest effect on the density of ceramics sintered at low temperatures.

Table 3. Properties of SPSed ceramic specimens from submicron Series No. 2 and No. 3 α-Al$_2$O$_3$ powders under continuous heating (Mode A). For Series No. 2 powders, temperatures $T_1$, $T_2$, $T_3$ are 1320, 1420, 1520 °C while for Series No. 3 powders they are 1470, 1530, 1600 °C.

| No. | T | $V_h$, °C/min | Series No. 2 | | | | Series No. 3 | | | |
|---|---|---|---|---|---|---|---|---|---|---|
| | | | d, μm | $\rho/\rho_{th}$, % | $H_v$, GPa | $K_{IC}$, MPa·m$^{1/2}$ | d, μm | $\rho/\rho_{th}$, % | $H_v$, GPa | $K_{IC}$, MPa·m$^{1/2}$ |
| 1 | | 10 | 0.5 | 99.46 | 22.3 | 2.2 | 2.1 | 97.98 | 18.4 | 2.7 |
| 2 | | 50 | 0.3 | 96.26 | 19.5 | 2.3 | 1.4 | 96.11 | 17.4 | 2.7 |
| 3 | $T_1$ | 100 | 0.2 | 94.75 | 20.0 | 2.6 | 1.0 | 92.22 | 15.7 | 3.4 |
| 4 | | 250 | 0.2 | 92.16 | 17.9 | 2.6 | 1.0 | 91.78 | 15.5 | 3.5 |
| 5 | | 350 | 0.2 | 92.56 | 18.2 | 2.5 | 1.0 | 92.10 | 15.6 | 3.1 |

| 6 | | 700 | 0.2 | 92.30 | 18.9 | 2.4 | 1.0 | 91.71 | 15.8 | 3.6 |
| 7 | | 10 | 1.7 | 99.66 | 20.0 | 2.4 | 12.9 | 98.25 | 16.2 | 2.0 |
| 8 | | 50 | 0.9 | 99.58 | 20.1 | 2.4 | 4.3 | 98.24 | 17.7 | 2.8 |
| 9 | $T_2$ | 100 | 0.9 | 99.52 | 19.8 | 2.3 | 2.7 | 97.39 | 17.1 | 2.6 |
| 10 | | 250 | 0.8 | 99.38 | 20.6 | 2.4 | 2.5 | 96.53 | 17.9 | 2.5 |
| 11 | | 350 | 0.7 | 99.06 | 20.3 | 2.1 | 2.2 | 96.38 | 18.1 | 2.6 |
| 12 | | 700 | 0.6 | 98.94 | 20.1 | 2.3 | 1.5 | 95.72 | 17.2 | 2.7 |
| 13 | | 10 | 5.1 | 99.72 | 18.6 | 2.5 | 20 | 98.24 | 16.1 | 1.7 |
| 14 | | 50 | 3.0 | 99.67 | 18.2 | 2.3 | 10.6 | 98.14 | 15.9 | 2.1 |
| 15 | $T_3$ | 100 | 2.8 | 99.60 | 17.9 | 2.5 | 8.0 | 98.00 | 16.7 | 2.7 |
| 16 | | 250 | 2.0 | 99.47 | 17.2 | 2.4 | 6.3 | 97.66 | 17.6 | 2.3 |
| 17 | | 350 | 1.9 | 99.47 | 16.9 | 2.1 | 6.1 | 97.52 | 15.9 | 2.7 |
| 18 | | 700 | 1.8 | 99.28 | 17.8 | 2.4 | 6.1 | 97.20 | 16.7 | 2.5 |

Fig. 11 shows the electron microscopy results for the alumina microstructure in specimens obtained by heating at different rates to 1320, 1420, and 1520 °C. Table 3 summarizes the results of the microstructural studies. An analysis of these results suggests that at $V_h$ = 10 °C/min, increasing the heating temperature from 1320 to 1520 °C expands the average grain size of alumina from 0.5 μm (Fig. 11a) to 5.1 μm (Fig. 11g). The specimens sintered at 1520°C have a uniform microstructure, without any anomalously large grains (Figs. 11g, h, k). Higher heating rates reduce the average grain size and cause finer-grained microstructure to be formed. As can be seen in Table 3, increasing $V_h$ from 10 to 700 °C/min decreases the average grain size from 5.1 to 1.8 μm (Figs. 10g, k). A similar heating rate increase to temperatures of 1320 and 1420 °C reduces the average grain size by 0.3 and 1.1 μm, respectively (Table 3). It can therefore be concluded that the ceramic grain microstructural parameters are more sensitive to changes in $V_h$ at high sintering temperatures. At low temperatures, nano- and submicron pores can act as stoppers for migrating grain boundaries [19] and this effect of $V_h$ is much less significant.

As an example, Fig. 12 demonstrates the results of TEM examination of the ceramics microstructure. It is evident that the specimens sintered at low temperatures have a uniform UFG microstructure and an increased volume fraction of nano- and submicron pores. The pores are predominantly located at grain boundaries triple junctions. Interestingly, even after high-temperature sintering, a large number of pores are faceted (Fig. 12c), which we believe to be an indirect indication of low diffusion intensity. The grain boundaries of the sintered ceramics are clean, with an ordered crystalline structure and without any excessive secondary or amorphous phase precipitates (Fig. 12d).

If sintered at low temperatures and low heating rates, ceramics demonstrate very good microhardness performance for pure alumina. Alumina sintered at a heating rate of 10 °C/min and at 1320 °C has a microhardness of $H_v = 22.3 \pm 1$ GPa (Table 3). Increasing the heating rate to 700 °C/min brings microhardness down to $18.9 \pm 1$ GPa. The minimum fracture toughness coefficient $K_{IC}$ varies insignificantly from 2.2 to 2.4-2.6 MPa·m$^{1/2}$, which is within its measuring error ($\pm 0.3$ MPa·m$^{1/2}$).

If the sintering temperature is raised to 1520°C, the ceramics microhardness deteriorates due to grain growth, providing, however, a healthy combination of microhardness and fracture toughness. If ceramics are sintered at 1520 °C at a heating rate of 10 °C/min, their microstructure becomes uniform and highly dense (99.72%) with a grain size of ~5 μm and their mechanical performance improves: $H_v = 18.6$ GPa, $K_{IC} = 2.5$ MPa·m$^{1/2}$.

Fig. 8b shows the dependences of the shrinkage of Series No. 2 submicron powders on isothermal holding time at various temperatures. Table 4 summarizes the results of researching the effect of isothermal holding time on ceramics density, average grain size and mechanical properties.

Table 4. Properties of SPS-processed ceramic specimens from Series No. 2-3 α-Al$_2$O$_3$ submicron powders under isothermal holding (Mode B). For Series No. 2 powders, temperatures T$_1$, T$_2$, T$_3$ are 1320, 1420, 1520 °C while for Series No. 3 powders they are 1470, 1530, 1600 °C.

| No. | T | t, min | Series No. 2 | | | | Series No. 3 | | | |
|---|---|---|---|---|---|---|---|---|---|---|
| | | | d, μm | ρ/ρ$_{th}$, % | H$_v$, GPa | K$_{IC}$, MPa·m$^{1/2}$ | d, μm | ρ/ρ$_{th}$, % | H$_v$, GPa | K$_{IC}$, MPa·m$^{1/2}$ |
| 1 | T$_1$ | 0 | 0.22 | 96.26 | 19.5 | 2.3 | 1.4 | 96.11 | 17.7 | 2.8 |
| 2 | | 3 | 1.7 | 99.61 | 20.3 | 2.4 | 2.6 | 97.42 | - | - |
| 3 | | 10 | 2.2 | 99.71 | 19.6 | 2.5 | 3.9 | 98.02 | - | - |
| 4 | | 30 | 2.9 | 99.71 | 19.0 | 2.8 | 5.2 | 98.10 | - | - |
| 5 | T$_2$ | 0 | 1.0 | 99.58 | 20.1 | 2.4 | 1.4 | 98.24 | 17.1 | 2.5 |
| 6 | | 3 | 4.3 | 99.64 | 18.5 | 2.6 | 6.1 | 97.93 | - | - |
| 7 | | 10 | 5.6 | 99.69 | 18.3 | 2.6 | 9.8 | 98.10 | - | - |
| 8 | | 30 | 7.6 | 99.72 | 17.3 | 2.9 | 12.7 | 98.26 | - | - |
| 9 | T$_3$ | 0 | 2.8 | 99.67 | 18.1 | 2.3 | 3.7 | 98.14 | 15.9 | 2.1 |
| 10 | | 3 | 10.9 | 99.65 | 16.9 | 3.2 | 9.4 | 98.07 | - | - |
| 11 | | 10 | 13.4 | 99.66 | 17.5 | 3.0 | 11.5 | 98.15 | - | - |
| 12 | | 30 | 16.5 | 99.69 | 16.0 | 3.1 | 22.6 | 98.19 | - | - |

Fig. 8b suggests that raising the isothermal holding temperature from 1320 to 1520°C increases the maximum powder shrinkage from ~0.25 to ~0.42 mm and the ceramic density from 99.71 to 99.69% (Table 4). As can be seen in Table 4, grains manifest intensive growth during isothermal exposure, decreasing the material microhardness and increasing fracture toughness. The highest values of the coefficient K$_{IC}$ = 3.1–3.2 MPa·m$^{1/2}$ can be achieved following isothermal holding over 30 min at 1520 °C. It should be noted that, after this exposure, the microstructure of the ceramic specimens contains quite large pores up to 8-10 μm long at the aluminum oxide grain boundaries (Fig. 13). Fig. 13b has these large pores marked with a yellow dotted line. It can be assumed that after isothermal holding under simultaneous pressure, large pores are formed due to coalescence or strain-induced growth of submicron pores.

### 3.3. Fine powder. Series No. 3

Research into the sintering kinetics of Series No. 3 aluminum oxide fine powders consisted in heating the powders to temperatures of 1470, 1530, 1600 °C at rates of 10-700 °C/min. Shrinkage (L) and shrinkage rate (S) dependences of Series No. 3 fine powders on heating temperature with different $V_h$ are shown in Figs. 10c, d. An analysis of L(T) and S(T) curves suggests that the effect of heating rates on the sintering kinetics of Series No. 3 fine powders is similar by nature to the effect of $V_h$ on L(T) and S(T) curves for powders of Series No. 1 (Fig. 6) and No. 2 (Figs. 10a, b). It should be noted that when fine powders are subjected to SPS, their $L_{max}$ and $S_{max}$ decrease more significantly as the heating rate $V_h$ grows (Figs. 10c, d). Curves L(T) and S(T) for Series No. 3 powders are shifted to the region of higher heating temperatures compared to L(T) and S(T) curves for submicron powders of Series No. 1 and 2. For example, fine powders with an initial particle size R ~ 1 μm have the shrinkage finishing temperature of 1450–1460 °C and 1620–1630 °C at $V_h$ = 10 and 700 °C/min, respectively. This is ~150–200 °C higher than the shrinkage finishing temperature of Series No. 2 submicron powders heated at 10 °C/min (shrinkage finishing temperature 1265–1280 °C) and 700 °C/min (shrinkage finishing temperature 1475–1500 °C). Pair comparison of Figs. 10a – 10c and Figs. 10b – 10d demonstrates that the maximum values of $L_{max}$ and $S_{max}$ for Series No. 3 fine powders are less than for Series No. 2 submicron powders sintered under similar conditions. At a heating rate of 10 °C/min, the maximum shrinkage rate for Series No. 2 powder is $S_{max}$ = 2.2·10$^{-3}$ mm/s while for Series No. 3 fine-grained powder, it is $S_{max}$ ~ 1.5·10$^{-3}$ mm/s. Increasing the heating rate to 50 and 100 °C/min further expands the gap between $S_{max}$ for powders in Series No. 2 and Series No. 3. At $V_h$ = 50 and 100 °C/min, $S_{max}$ for Series No. 2 submicron powder reaches 7.1·10$^{-3}$ and

13.4·10⁻³ mm/s, respectively. When Series No. 3 powder is sintered at similar rates, $S_{max}$ is 4.7·10⁻³ mm/s ($V_h$ = 50 °C/min) and 8.3·10⁻³ mm/s ($V_h$ = 100 °C/min). Thus, increasing the initial size of $Al_2O_3$ particles expectedly resulted in lower shrinkage rates. This finding is well aligned with the literature data [52], also correlating with the density measurement results for ceramics that are produced at various heating rates. Comparison of the data in Table 3 suggests that ceramics sintered from Series No. 3 fine powder have a lower relative density than ceramics sintered from Series No. 2 powder under similar heating conditions.

As can be seen in Table 3, higher SPS heating rates reduce the density of $Al_2O_3$ ceramics. Similarly to Series No. 2 powders, the density reduction under increased heating rates manifests most forcefully at low sintering temperatures. For ceramics sintered from Series No. 3 powder at 1470 °C increasing the heating rate from 10 to 700 °C/min reduces density from 97.9 to 91.7% ($\Delta\rho$ = 6.2%) while for ceramics sintered at T = 1600 °C a similar increase in $V_h$ decreases density from 98.2% to 97.2% ($\Delta\rho$ = 1%).

Fig. 14 shows electron microscopic images of fractures in alumina specimens sintered at various temperatures and heating rates. These images suggest that a fairly uniform grain microstructure is formed in the ceramic at all heating rates and sintering temperatures. No traces of abnormal grain growth were detected. It should be noted that the ceramic microstructure has quite clearly visible submicron pores that are located along the grain boundaries and in the bulk of aluminum oxide grains. Raising the heating rate increases the volume fraction of pores, which correlates with the density measurements results (Table 3).

TEM confirms that pores are formed at high heating rates (Fig. 15). These figures demonstrate that raising heating rates increases the number of submicron pores that are located within the crystal lattice and at grain boundaries of aluminum oxide. Most of the pores are of a nearly equiaxed shape.

Table 3 shows that increasing the heating rate decreases the average grain size. Heating at $V_h$ = 10 °C/min to 1470 °C forms a microstructure with a grain size of 2.1 μm but heating at a rate of 700 °C/min to the same temperature can reduce the average grain size of alumina ~ 1 μm. Thus, increasing the heating rate from 10 to 700 °C/min can reduce the average grain size of the alumina sintered at T = 1470 °C by Δd = 1.1 μm. A similar effect can be observed when heating specimens to 1530 and 1600 °C but the average grain size decreases even more – Δd = 11.4 μm at T = 1530 °C; Δd = 13.9 μm at T = 1600 °C (Table 3).

The microhardness of fine-grained alumina sintered from Series No. 3 powder turns out less than for ceramics sintered under the same conditions from Series No. 2 powder (see Table 3). It should be noted that ceramics with the same grain size (d ~ 2 μm), but made from different powders, vary in microhardness: for ceramics sintered from Series No. 3 fine powder, the microhardness is $H_v$ = 18.4 GPa, which is 10% lower than for ceramics sintered from Series No. 2 submicron powder ($H_v$ = 20 GPa) (see Table 3).

An analysis of the data provided in Table 3 suggests that reducing the ceramic density decreases its microhardness while increasing the minimum fracture toughness coefficient. For example, raising the heating rate from 10 to 700 °C/min decreases the ceramic density from 97.98% to 91.71% and reduces $H_v$ from 18.4 to 15.8 GPa. At the same time, fracture toughness grows from 2.7 to 3.6 MPa·m$^{1/2}$. An increase in the average grain size (raising the sintering temperature) also decreases microhardness, yet the fracture toughness coefficient changes rather insignificantly in this case (Table 3).

An analysis of the findings of research into the microstructure and properties of ceramics sintered in Mode B can be found in Table 4 and suggest that increasing the isothermal holding time increases the ceramics density. It should be also noted that the relative density of ceramics sintered at 1600°C is the same as the density of ceramics obtained by holding for 30 minutes at 1530°C (Table 4). Thus, raising the holding temperature from 1530 to 1600°C did not entail the expected increase in the ceramics density. We assume this to be due to the formation of large pores along the boundaries of alumina grains (Fig. 16). It should be also noted that holding at 1530 and 1600°C is accompanied by anomalous grain growth when the first anomalously large grains of ~50 µm in size are detected after holding at 1530°C for 3 min (Fig. 16a). Raising the exposure temperature or time increases the size of abnormally large grains, the average size of which reaches ~150 µm at 1600 °C (Fig. 16). The volume fraction of anomalously large grains is not very large, however, and the average grain size calculated by the chord method is ~22.6 µm (Table 4).

## 4. Discussion

### 4.1. Grain growth analysis

Grain growth in alumina under isothermal holding (Mode B) will be analyzed below. As can be seen in Table 4 and Fig. 17, after sintering Series No. 2 powders in the isothermal holding mode, grains begin to grow abnormally fast. Sintering Series No. 2 powder for 30 min at 1320°C creates a high-density microstructure with a grain size of 2.9 µm (Table 4). A similar holding of equally fine Series No. 1 powder forms a homogeneous UFG microstructure with a grain size of ~ 0.5 µm. Ceramics sintered from Series No. 2 powder at 1520°C for 30 min has an average grain size of ~16.5 µm whereas the ceramics sintered from a coarser Series No. 3 powder has an average grain size ~12.7 µm after holding for 30 minutes at 1530°C (Table 4).

Thus, the grain growth rate during sintering of ceramics from Series No. 2 powder is several times higher than the grain growth rate observed during sintering of equally fine Series No. 1 powder. This is a highly unusual behavior. It should be also noted that the average grain size in ceramics sintered at 1520 °C from Series No. 2 powder ($R_0$ ~ 0.2 µm) turns out larger than the average grain size in ceramics SPSed at 1530°C from Series No. 3 powder ($R_0$ ~ 1 µm) (Fig. 17).

Fig. 17 demonstrates that in ln(d) – ln(t) logarithmic coordinates, d(t) dependences can be interpolated with good accuracy by straight lines; the lineralization validity factor is $R^2 > 0.93$, except for ln(d) – ln(t) curve for Series No. 1. In the case of Series No. 1 and No. 3 ceramics, the slope of ln(d) – ln(t) curves is close to ~ 0.3, which corresponds to the case of power-law grain growth $d^3 - d_0^3 = At$ that is often observed in pure aluminum oxide [52]. For Series No. 2 ceramics, the slope of ln(d) – ln(t) curves corresponds to the case $d^4 - d_0^4 = At$. Exponential grain growth with a factor k = 4 can also be observed when sintering pure alumina [14].

The main difference between Series No. 2 and Series No. 1 powders is the amorphous layer on the surface of aluminum oxide particles (Fig. 3c). It should be noted that there is no amorphous layer at the grain boundaries after sintering (Fig. 12d). This may suggest that heating resulted in transforming the amorphous structure into one of the crystalline modifications of aluminum oxide. As the bulk of an amorphous material contains an excess free volume [53-55], then the "amorphous phase → crystalline structure" transformation creates an excessive density of vacancy and/or dislocation defects at the grain boundary. As the theory of nonequilibrium grain boundaries [55] demonstrates, this type of defects at grain boundaries increases the coefficient of grain boundary diffusion and, consequently, their migration mobility.

### 4.2. Analysis of Mechanical Properties

Fig. 18a shows a microhardness ($H_v$) – relative density ($\rho/\rho_{th}$) diagram generalizing the findings of the research into the alumina mechanical properties shown in Tables 2–4. As can be seen in Fig. 18a, the highest ceramic microhardness (more than 20 GPa) can be detected for a small (~0.5 μm) grain size and high relative density (more than 99%). It is interesting to note that for specimens sintered from Series No. 1 submicron powder under isothermal holding, a close-to-linear dependence between microhardness and porosity is observed (the linearalization validity factor is 0.9905).

The dependence of the alumina microhardness on the grain size $H_v(d)$ shown in Fig. 18b has a pronounced two-stage nature with a maximum. For ceramics sintered from submicron powders of Series No. 1 and No. 2, the maximum microhardness is reached at an average grain size $d_{max}$ ~ 0.5 μm while for alumina sintered from Series No. 3 fine powders, the maximum microhardness can be reached at an average grain size $d_{max}$ ~ 2.1–2.2 μm. An analysis of the findings in Tables 2–4 suggests that the decrease in microhardness in the $d < d_{max}$ region is associated with a lower ceramic density caused by low sintering temperature and/or high heating rate. Reduced microhardness in the $d > d_{max}$ region is well-known (see [56, 57]).

Fig. 18b demonstrates that, at the same grain size, the microhardness of ultrafine-grained ceramics with a grain size of less than 1 μm that have been sintered from Series No. 2 powder is on average 1–1.5 GPa higher than the microhardness of ceramics sintered from Series No. 1 powders. At $d = d_{max}$, the microhardness of Series No. 1 and 2 ceramics are practically the same within the measuring error. We reckon this to be due to a higher porosity of the alumina specimens sintered from Series No. 1 powders. As comparison of the data in Table 2 and Table 3 suggests, ceramic specimens of Series No. 1 have a lower relative density than Series No. 2 ceramics. Specifically, a Series No. 1 ceramic with a grain size of 0.3 μm has a relative density of $\rho/\rho_{th} = 94.5\%$ ($H_v = 18.6$ GPa) while a Series No 2 ceramic with the same grain size has a relative density of $\rho/\rho_{th} = 96.26\%$ and therefore a higher microhardness $H_v = 19.5$ GPa. With an average grain size of $d = 0.2$ μm, Series No. 2 ceramics have a density of 92.30–94.75% and $H_v = 18.2–20$ GPa while Series No. 1 ceramics have a density of 90.5–94.1 and $H_v = 16.1–18.6$ GPa (Tables 2–4).

Another factor contributing to the increased microhardness of ceramics sintered from Series No. 2 powder may be down to dislocation defects at the grain boundaries, which were formed during the crystallization of a thin amorphous surface layer during SPS (see section 4.1). At high sintering temperatures, the defect density at the grain boundaries decreases and the differences between the microhardness of ceramics sintered from powders of Series No. 1 and No. 2 become insignificant.

### 4.3. Analysis of sintering processes

The dependences of the aluminum oxide powders shrinkage on the heating temperature (Figs. 6, 10) are of a conventional three-stage nature (see [52]). This suggests that what occurs when alumina ceramics are subjected to SPS can be described as a sequence of processes of the initial (I), intermediate (II), and final (III) stages of sintering [52].

As per [52], the initial stage of sintering (Stage I) is characterized by forming contacts between powder particles. The transition between Stages I and II occurs at $\rho/\rho_{th} \sim$ 65-70%. Stage II is characterized by a larger area of contacts between particles and by intense powder compaction. It is important to point out that there is virtually no grain growth in ceramics at Stage II. At Stage III ($\rho/\rho_{th} >$ 90%), occluded porosity is formed, isolated pores undergo diffusion-controlled dissolution, and grains grow.

The Young–Cutler model [58] can be used to describe the kinetics of powder sintering at the beginning of sintering (Stage I). The model describes the initial stage of nonisothermal sintering of spherical particles under simultaneous lattice and grain boundary diffusion processes, as well as under creep. In the Young–Cutler model, the dependence between relative shrinkage ($\varepsilon$) and continuous heating temperature can be described by the following formula:

$$\varepsilon^2 \left(\frac{\partial \varepsilon}{\partial t}\right) = \left(\frac{2{,}63\gamma\Omega D_v \varepsilon}{kTd^3}\right) + \left(\frac{0{,}7\gamma\Omega bD_b}{kTd^4}\right) + \left(\frac{Ap\varepsilon^2 D}{kT}\right) \tag{1}$$

where $\gamma$ is the free surface energy, $D_v$ is the lattice (volume) diffusion factor, $D_b$ is the grain boundary diffusion factor, $D$ is diffusion factor under creep, $k$ is the Boltzmann constant.

As per [58], the slope of the dependence between shrinkage and temperature in $\ln(T \cdot \partial\varepsilon/\partial T) - T_m/T$ coordinates corresponds to the effective sintering activation energy $mQ_{s2}$, where $m$ is the factor that is dependent on the dominant sintering mechanism ($m = 1/3$ – for grain boundary diffusion, $m = 1/2$ – for lattice diffusion, $m = 1$ for viscous flow (creep)). When analyzing the findings, the melting temperature $T_m$ was assumed to be 2326 K. Fig. 19 shows, as an example, $\ln(T \cdot \partial\varepsilon/\partial T) - T_m/T$ dependences for Series No. 1 submicron powders.

$\ln(T \cdot \partial\varepsilon/\partial T) - T_m/T$ dependences are of a conventional three-stage nature (see [58]). At the intense compaction stage, $\ln(T \cdot \partial\varepsilon/\partial T) - T_m/T$ dependence can be interpolated fairly accurately with a straight line (Fig. 19a). When the activation energy $mQ_{s2}$ was being determined, a portion of $\ln(T \cdot \partial \varepsilon / \partial T) - T_m/T$ dependence was selected where the lineralization validity factor was at least 0.9 (Fig. 19a). If the lineralization validity of at least 0.9 could not be ensured, then such findings were disregarded (see, e. g., $\ln(T \cdot \partial \varepsilon / \partial T) - T_m/T$ dependences in Fig. 19b for heating rates of 10 and 50 °C/min).

As can be seen in Fig. 19b, Series No. 1 submicron powders at $V_h$ = 100–350 °C/min have the effective activation energy $mQ_{s2}$ of 8.4-9.2 $kT_m$. At the representative value $m$ = 1/3, which is typical of SPS of ultrafine-grained ceramics [33, 34, 59], the activation energy of sintering $Q_{s2}$ Series No. 1 submicron powders is 25.2-27.6 $kT_m$ (~ 487-533 kJ/mol). This activation energy turns out slightly higher than the activation energy of the grain boundary diffusion of oxygen ions in aluminum oxide ($Q_b$ ~ 380 kJ/mol ~ 19.7 $kT_m$ [60]) and is intermediately placed between $Q_b$ and the activation energy of crystal lattice diffusion of oxygen ions ($Q_v$ ~ 636 kJ/mol [60-62]).

The effective activation energy $mQ_{s2}$ for SPS of Series No. 2 submicron powders lies in the range from ~7.2 $kT_m$ at $V_h$ = 10 °C/min to 8.2 $kT_m$ at $V_h$ = 350 °C/min. The number of points is low on $\ln(T \cdot \partial \varepsilon / \partial T) - T_m/T$ dependence at $V_h$ = 700 °C/min; the lineralization validity factor is low (~0.71) and therefore these findings were discounted in subsequent analysis. At $m$ = 1/3, the activation energy of SPS of Series No. 2 submicron powders is ~ 21.6-24.6 $kT_m$ (~ 418-475 kJ/mol). These $Q_{s2}$ values for Series No. 2 powders prove to be 10% lower than the SPS activation energy of equally fine Series No. 1 powders and are close to the activation energy of grain boundary diffusion $Q_b$. We believe that the reason for lower SPS activation energy consists in the excessive defect density at grain boundaries of ceramics sintered from Series No. 2 powders. It should be noted that at lower heating temperatures, an amorphous layer at the intense shrinkage stage can facilitate the sintering of submicron $Al_2O_3$ particles.

The $mQ_{s2}$ for Series No. 3 fine aluminum oxide powders varies from 10.9 to 14.5 $kT_m$. At $m$ = 1/3, the activation energy $Q_{s2}$ turn out to be abnormally high, which is not in line with the known values of the activation energies for aluminum oxide diffusion. At $m$ = 1/2, the activation energy of SPS of Series No. 3 submicron powders is ~ 21.8-29 $kT_m$ (~ 421-560 kJ/mol). This $Q_s$ value lies intermediately between the activation energy of oxygen ion diffusion along grain boundaries ($Q_b$ ~ 380 kJ/mol [60]) and the activation energy of oxygen ions lattice diffusion ($Q_v$ ~ 636 kJ/mol [60-62]). This implies that the intensity of sintering Series No. 3 fine-grained powders must be limited by crystal lattice and grain boundary diffusion occurring in parallel. The enhanced contribution of diffusion in the crystal lattice during SPS of fine powders must be due to their higher sintering temperatures as well as a large initial particle size and, therefore, a coarser-grained structure of the sintered ceramics.

Thus, an intermediate conclusion can be made that kinetics of Series No. 1 and No. 3 powders sintering at the intermediate stage of intensive compaction (Stage II) must be determined by grain boundary and crystal lattice diffusion occurring in parallel. Series No. 2 powders sintering kinetics at the stage of intensive compaction is determined by the intensity of grain boundary diffusion.

Fig. 18a demonstrates that at higher sintering temperatures, the slope of $\ln(T \cdot \partial \varepsilon / \partial T) - T_m/T$ dependence becomes negative, which is typical of analyses of L(T) shrinkage curves using the Young–Cutler model [58] (see also [33, 34, 59]). The Coble method [52, 63], which has been redesigned for continuous heating, will be used to determine the SPS mechanism in the region of elevated heating temperatures.

Let us assume that at elevated temperatures during Stage III of sintering (with $\rho/\rho_{th}$ > 90% [52, 63, 64]), the rate of change in relative shrinkage $\dot{\varepsilon}$ is proportional to the strain rate for the porous body. If the temperature dependence L(T) is known, the duration of intense shrinkage stage ($t_1, t_2, t_3, \ldots t_n$), corresponding to each average grain size ($d_1, d_2, d_3, \ldots d_n$) and relative density ($\rho_1, \rho_2, \rho_3 \ldots \rho_n$) in Tables 2 and 3, can be determined. This enables mapping between ($d_1, d_2, d_3, \ldots d_n$) data set and $\dot{\rho}/\rho$ compaction rates. As per [52], at the final stage of hot compacting, $\dot{\rho}/\rho(d)$ dependence when bulk or grain boundary diffusion prevails can be described with equations (2) и (3), respectively:

$$\dot{\rho}/\rho = \frac{40}{3}(D_v/kT)(\Omega/d^2)(p_a\phi + 2\gamma_{sv}/r), \tag{2}$$

$$\dot{\rho}/\rho = \frac{15}{2}(\delta D_b/kT)(\Omega/d^3)(p_a\phi + 2\gamma_{sv}/r). \tag{3}$$

where $\delta$ is the grain boundary width, $\Omega$ is the atomic volume, $p_a$ is stress applied externally, $\phi$ is the stress concentration factor, $\gamma_{sv}$ is the solid-gas interface energy, $r$ is the pore radius.

As can be seen in equations (2) и (3), the slope of $\ln(\dot{\rho}/\rho) - \ln(d)$ dependence is driven by the dominant diffusion mechanism. At a constant temperature, the slope of $\ln(\dot{\rho}/\rho) - \ln(d)$ dependence for lattice diffusion corresponds to $n = 2$ (with $D_v$ = const) and to $n = 3$ (with $D_b$ = const) for grain boundary diffusion [52].

Fig. 19 demonstrates $\ln(\dot{\rho}/\rho) - \ln(d)$ dependences that are calculated based on analysis of the data in Tables 3-4 for Series No. 2 and Series No. 3. Figs. 19a, b present the results of analyzing the data obtained for continuous heating (Mode A) and Figs. 19c, d show the findings for isothermal holding (Mode B).

Mode A results for Series No. 1 powders were not analyzed, as heating was performed at different temperatures (see Table 2) and therefore the condition of $D_v$ = const or $D_b$ = const was not met. The *n* factor for Series No. 1 under isothermal holding (Mode B) is $n \sim 2.6$ yet it should be noted that the lineralization validity factor is fairly low ($R^2 \sim 0.83 < 0.90$). An assumption can be made that, as per [52], grain boundary diffusion is dominant at the final SPS stage for Series No. 1 powder.

This conclusion is indirectly confirmed by the calculation of the SPS activation energy at the final sintering stage (Stage III). The data that were obtained for isothermal holding (Mode B) will be used to estimate the SPS activation energy. It follows from formula (3) that at d = const, the SPS activation energy ($Q_{s3}$) can be determined from the slope of $\ln(\dot{\rho}/\rho) - T_m/T$ dependence. Our analysis suggests that the sintering activation energy at Stage III is ~26 $kT_m$ (Table 5) for low heating temperatures (1390-1460 °C) where the ceramic microstructure remains stable. This is close to the sintering activation energy at stage II that is calculated as per the Young–Cutler model (24.6-27.6 $kT_m$).

Table 5. Analysis of SPS mechanisms for aluminum oxide powders of different degrees of fineness

| Series No. | Young–Cutler model (Stage II) | | | Coble model (Stage III) | |
|---|---|---|---|---|---|
| | $mQ_{s2}$, $kT_m$ | $m$ | $Q_{s2}$, $kT_m$ | $n$ | $Q_{s3}$, $kT_m$ |
| 1 | 8.4-9.2 | 1/3 | 25.2-27.6 | B: ~2.6[(1)] | ~26 |
| 2 | 7.2-8.2 | 1/3 | 21.6-24.6 | A: 3.5-3.8<br>B: 4.0-5.6 | - |
| 3 | 10.9-14.5 | 1/2 | 21.8-29.0 | A: 2.6-3.5<br>B: 2.6-3.3 | 22.8-25.0 |

Note: [(1)] – with $R^2 = 0.82$-$0.83$

The slope of $\ln(\dot{\rho}/\rho) - \ln(d)$ dependence for Series No. 2 powders in Mode A lies in the range from ~3.5 to 3.9 (Fig. 19a) while in the case of isothermal holding sintering (Mode B), it is $n \sim 4.0$-$5.6$ (Fig. 19b). We were unable to properly determine the SPS activation energy at Stage III because there were no data for Mode B that would meet the criterion of $d_1 = d_2 = d_3 = $ const for various isothermal holding temperatures $T_1$, $T_2$, $T_3$ (see Table 4).

Attention should be drawn to abnormally high $n$ factor in Coble equation for Series No. 2 powders tested in Mode B. The authors are as yet unable to provide a definite answer to the question about the reasons for this high value of the $n$ factor in the Coble equation for isothermal holding testing. It can be assumed that one of the probable reasons for the higher $n$ factor consists in an increased defect density at the grain boundaries of Series No. 2 ceramics. Since the defect concentration at the aluminum oxide grain boundaries decreases as the temperature and sintering times grow, the diffusion factor during sintering becomes variable ($D_b \neq$ const). This fails the $D_b$ = const condition, which is necessary for correctly evaluating the *n* factor using equation (3).

An analysis of SPS data for Series No. 3 powders demonstrates that the slope of $\ln(\dot{\rho}/\rho)$ – ln(d) is close to the theoretical value of factor *n* ~ 3 (Table 5), which corresponds to the case of grain boundary diffusion. This suggests that at Stage III, in the region of elevated temperatures, the intensity of sintering fine aluminum oxide powders is limited by the intensity of grain boundary diffusion. The SPS activation energy at Stage III is ~ 22.8-25 $kT_m$, which is slightly higher than $Q_b$ ~ 380 kJ/mol ~ 19.7 $kT_m$ [60]. It should be noted that in the case of creep higher activation energies, $Q_s$ are observed that exceed their theoretical values for the corresponding strain mechanism [65, 66]. This is often associated with internal stresses being formed during creep in fine-grained materials, leading to a threshold creep stress $\sigma_{cr}$ [67]. The $\sigma_{cr}$, threshold stress depends on the creep testing conditions and microstructural parameters and is normally not taken into account in equations (2) – (3) that are well-tested for coarse-grained materials.

For SPS of Series No. 3 fine powders, an increased density of dislocations in the initial powders (Fig. 4c) can be an extra factor that contributes to the formation of grain boundary defects. As images in Fig. 14 suggest, there are no dislocations in the grains of sintered ceramics. This can lead to the assumption that lattice dislocations hit the migrating grain boundaries during recrystallization (see also [55]). It can generate long-range internal stresses from grain boundaries [55] and reduce the grain boundary diffusion factor [55]. We believe that the higher diffusion permeability of the alumina migrating grain boundaries may be one of the factors contributing to the change of the dominant diffusion mechanism from diffusion in the crystal lattice to grain boundary diffusion during the transition from Stage II to Stage III.

It should be noted that internal stresses during SPS can explain, in our opinion, the pores in triple junctions of grain boundaries during aluminum oxide sintering (Figs. 12b, 15). Intensive grain growth in Series No. 2 and No. 3 ceramics increases the characteristic scale of the diffusion path (mass transfer) and hinders relaxation of internal stresses that arise from defects at the ceramic grain boundaries.

**Conclusions**

1. The spark plasma sintering (SPS) of submicron and fine aluminum oxide powders has a two-stage nature. An analysis of shrinkage curves using the Young–Cutler model demonstrates that at the intense compaction stage, the sintering kinetic of submicron $Al_2O_3$ powders is determined by the intensity of grain boundary diffusion while for fine powders with the initial particle size of ~1 μm, it is driven by simultaneous diffusion processes in the crystal lattice and along grain boundaries.

An amorphous layer on the surface of submicron $Al_2O_3$ particles (Series No. 2 powders) reduces the SPS activation energy and seems to facilitate sliding of the particles relative to each other during hot compaction.

2. The amorphous layer on the particle surface causes an abnormally fast grain growth at low temperatures where the average grain size in Series No. 2 ceramics becomes several times larger than the average grain size in ceramics sintered under the same conditions from submicron powders without an amorphous layer on the particle surface. The kinetics of grain growth in ceramics sintered from Series No. 2 submicron powder is described by the power law with $k$ in equation $d^k – d_o^k = At$ equaling $k = 4$. The $k$ factor in case of grain growth in ceramics sintered from submicron and micron $Al_2O_3$ powders is $k = 3$.

It has been suggested that an excessive density of vacancy and/or dislocation defects at grain boundaries can be the reason for the abnormally fast grain growth in Series No. 2 ceramics. Defects begin to occur during SPS due to crystallization of the amorphous layer that has an excessive free volume.

3. The Coble model has been used to demonstrate that in the region of elevated temperatures, the ceramic compaction intensity is limited by the intensity of grain boundary diffusion, irrespective of the initial particle size. It has been demonstrated that abnormally high values of $n$ ~4-5 in the Coble equation are observed during sintering of Series No. 2 powders. The values of the $n$ factor in the Coble equation for submicron and micron $Al_2O_3$ powders without an amorphous layer on their surface are close to their theoretical value of $n \sim 3$, corresponding to the case of grain boundary diffusion.

It has been suggested that the abnormally high $n$ for Series No. 2 ceramics is associated with the influence of grain boundary defects on the grain boundary diffusion factor as well as on internal stresses in the ceramic.

4. The dependence of microhardness on the grain size has a two-stage nature with a maximum. When ceramics are sintered from submicron powders, the maximum hardness is achieved with the average grain size of ~0.5 μm and with d = 2–2.5 μm when ceramics are sintered from fine powders. It has been demonstrated that with the grain size being the same, ceramics sintered from Series No. 2 powder have higher microhardness than ceramics sintered from the conventional submicron $Al_2O_3$ powder. It is assumed that higher microhardness in Series No. 2 ceramics are due to defects at the grain boundaries.

**Conflict of interest**. The authors declare that they have no known competing financial interests or personal relationships that could have appeared to influence the work reported in this paper.

**Acknowledgements**

The study was performed with the support of the Russian Science Foundation (Grant No. 20-73-10113).

TEM study of the microstructure was carried out on the equipment of the Center Collective Use "Materials Science and Metallurgy" with the financial support of the Ministry of Science and Higher Education of the Russian Federation (Grant No. 075-15-2021-696).

## List of Figures

Figure 1. Ceramic SPS diagrams: (a) temperature (T) and pressure (P) dependences on heating times; (b) temperature dependences of Series No. 3 powders shrinkage

Figure 2. Electron microscopic image (a, b, c) of Series No. 1 powder and XRD phase analysis results for powders in Series No. 1-3 (d)

Figure 3. Electron microscopic images of Series No. 2 powders: (a) SEM, (b, c, d) TEM

Figure 4. Results of XPS analysis of the powders for Series No. 2: overview spectra from the initial surface (a), after 30 s sputtering (b), and after 30 min sputtering (c); Al2p spectra from them the initial surface (d) and after 6 min sputtering (e); O1s spectra from the initial surface (f) and after 6 min sputtering (k)

Figure 5. Electron microscopic image of Series No. 3 powder: (a) SEM, (b, c) TEM

Figure 6. Graphs of dependences of shrinkage L (a) and shrinkage rate S (b) of Series No. 1 powders on the temperature of heating performed at different rates. Mode A

Figure 7. Fracture microstructure of ceramic specimens made from Series No. 1 powder by SPS, with heating at a rate of 10 (a), 50 (b), 250 °C/min (c). Mode A. SEM

Figure 8. Graph of shrinkage dependence on isothermal holding time for powders in Series No. 1 (a) and Series No. 2 (b). The holding temperature was 1350 °C for Series No. 1 powders while 1320, 1420, 1520 °C were set for Series No. 2 powders

Figure 9. Fracture microstructure of ceramic specimens SPSed from Series No. 1 powders in isothermal holding mode (Mode B) over 0 (a), 15 (b), 30 min (c). SEM

Figure 10. Dependences of shrinkage (a, c) and shrinkage rate (b, d) on the heating temperature of Series No. 2 powder (a, b) and Series No. 3 fine powder (c, d) at different rates. Mode A

Figure 11. Microstructure of fractures in alumina specimens obtained by heating Series No. 2 powders to 1320 °C (a, b, c), 1420 °C (d, e, f), and 1520 °C (g, h, k). Heating rate $V_h$ = 10 °C/min (a, d, g), 250 °C/min (b, e, h) and 700 °C/min (c, f, k). SEM

Figure 12. Microstructure of alumina SPSed specimens from Series No. 2 powder under the following modes: (a, d) $V_h$ = 350 °C/min, T = 1150 °C; (b) $V_h$ = 700 °C/min, T = 1150 °C, (c) $V_h$ = 700 °C/min, T = 1300 °C. TEM

Figure 13. Microstructure of ceramics produced in Mode B from Series No. 2 powder by isothermal holding for 15 min (a) and 30 min (b) at 1520 °C

Figure 14. Microstructure of fractures in alumina specimens obtained by heating Series No. 3 powders to 1470 °C (a, b, c), 1530 °C (d, e, f), 1600 °C (g, h, k). Heating rate $V_h$ = 10 °C/min (a, d, g), 250 °C/min (b, e, h), 700 °C/min (c, f, k). SEM

Figure 15. Microstructure of alumina SPSed specimens from Series No. 3 powder under the following modes: (a, d) $V_h$ = 10 °C/min, T = 1320 °C; (b) $V_h$ = 300 °C/min, T = 1320 °C, (c) $V_h$ = 700 °C/min, T = 1460 °C. TEM

Figure 16. Microstructure of ceramics with areas of abnormally large grains. Series No. 3. Isothermal holding for 3 min (a) and 30 min (b, c) at 1530 °C (a, b) and 1600 °C (c). Mode B

Figure 17. Dependences of alumina average grain size on isothermal holding times at different temperatures in logarithmic coordinates

Figure 18. Analysis of alumina mechanical properties measurements: (a) dependence of microhardness on relative density; (b) dependence of microhardness on grain size. The markers in Fig. 18 denote average ceramic grain size

Figure 19. $\ln(T \cdot \partial\varepsilon/\partial T) - T_m/T$ curves for Series No. 1 powders: (a) general representation of the dependence for heating rate $V_h = 100$ °C/min; (b) linear sections of $\ln(T \cdot \partial\varepsilon/\partial T) - T_m/T$ curves that are used to analyze the SPS activation energy

Figure 20. Dependence of density change rate on grain size in logarithmic coordinates: (a, b) Series No. 2 powders; (c, d) Series No. 3 powders; (a, c) Mode A; (b, d) Mode B

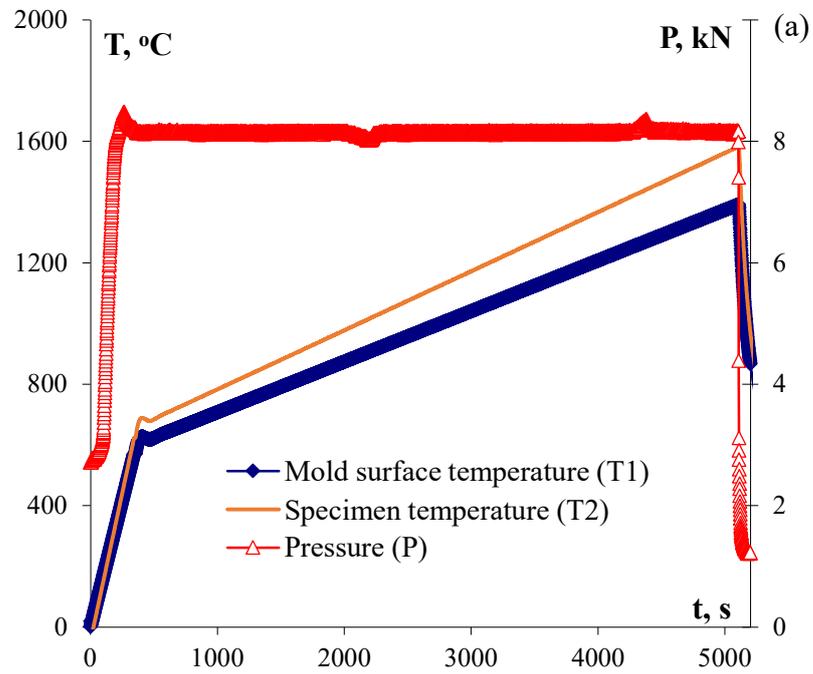

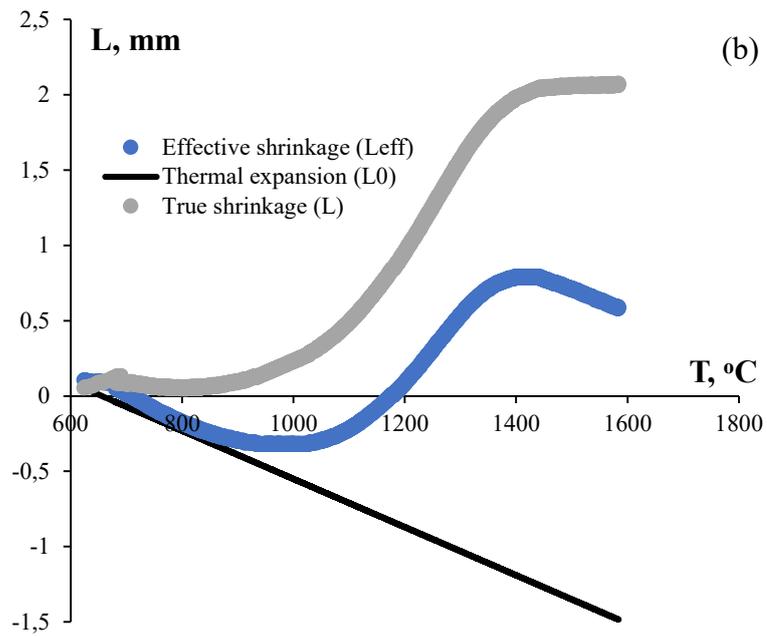

Figure 1

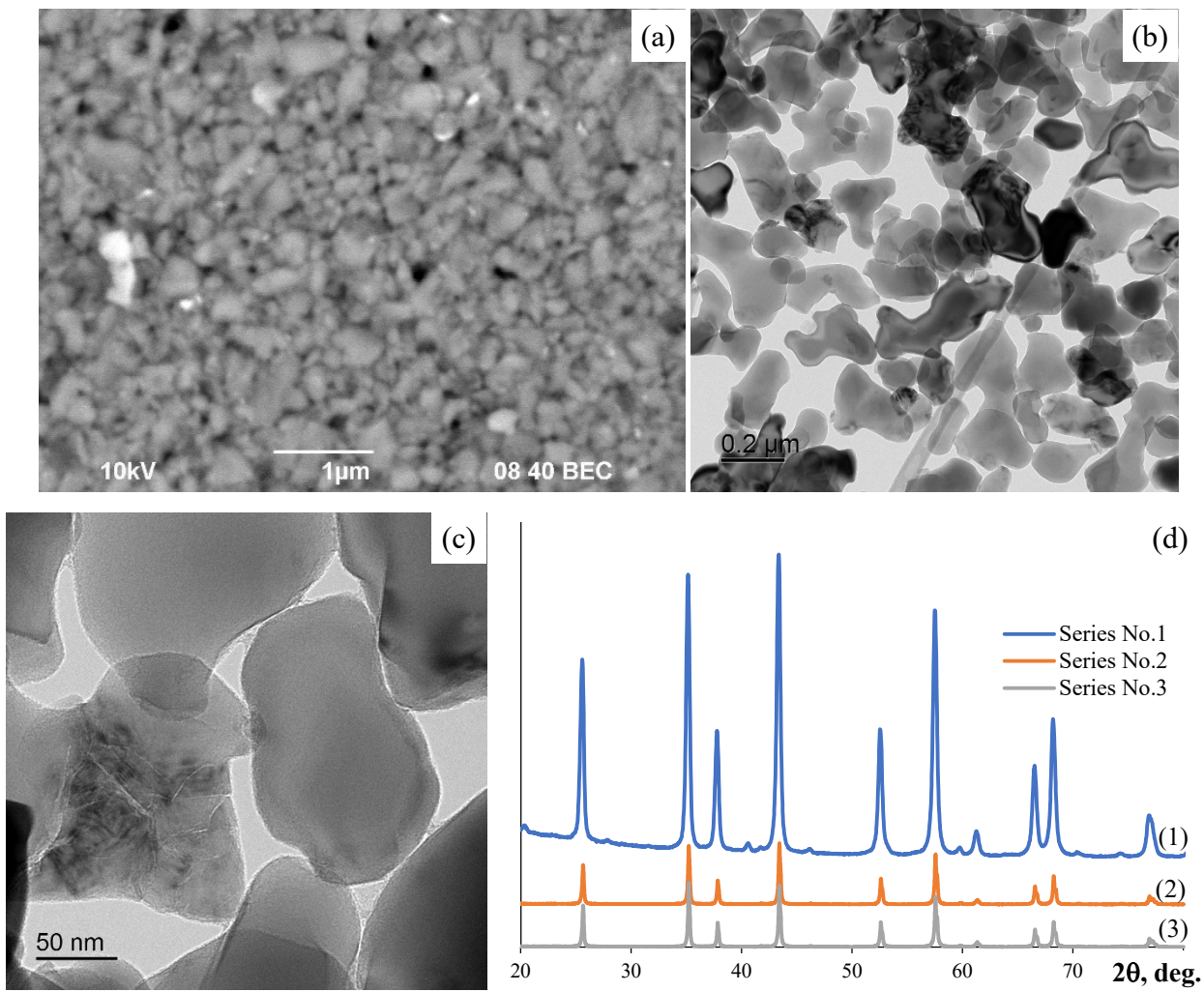

Figure 2

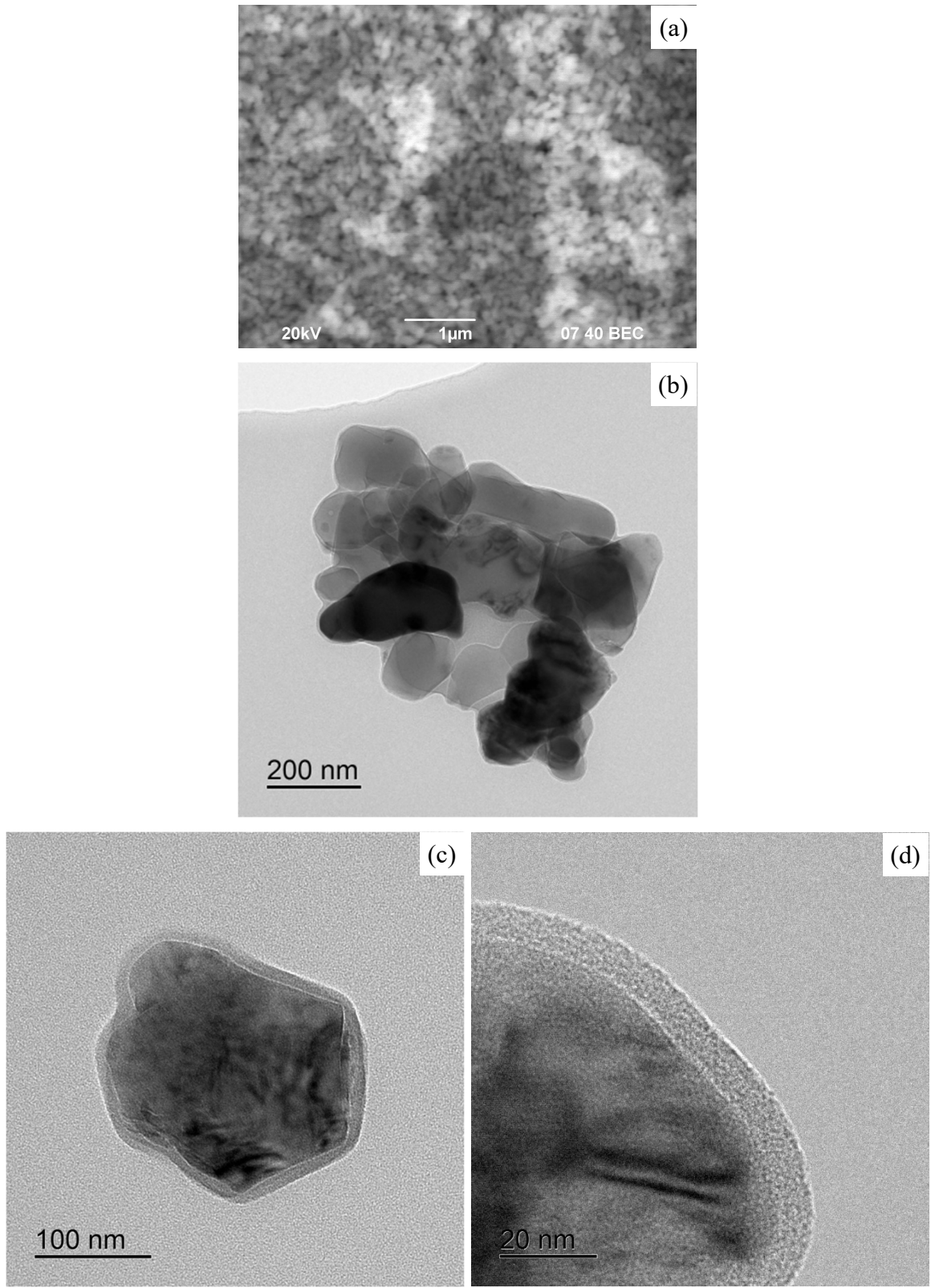

Figure 3

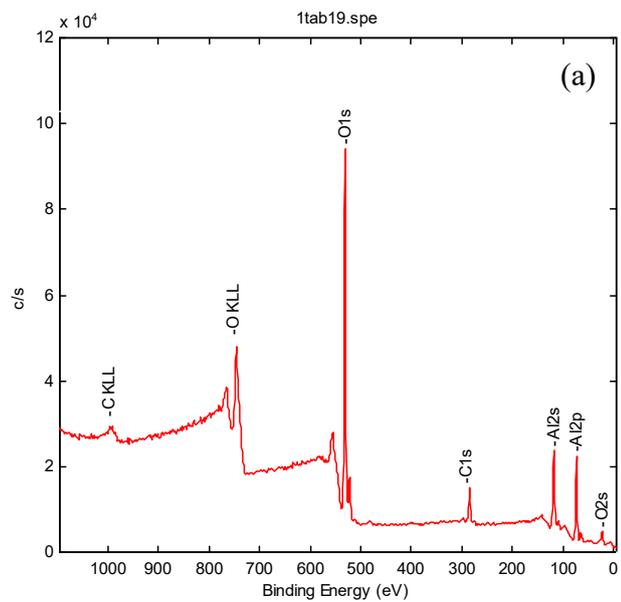
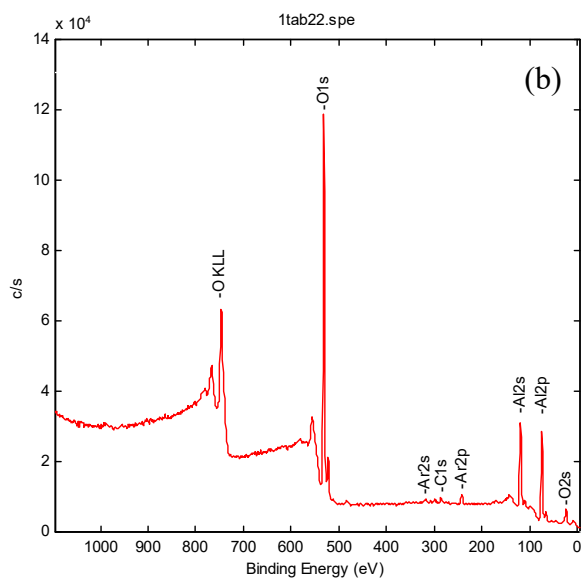
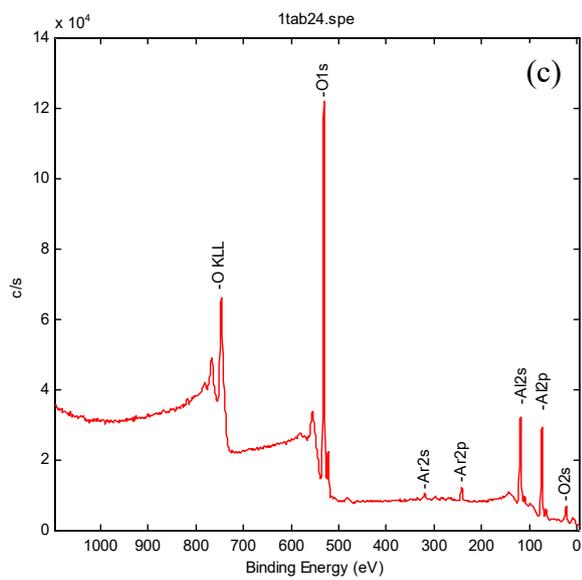

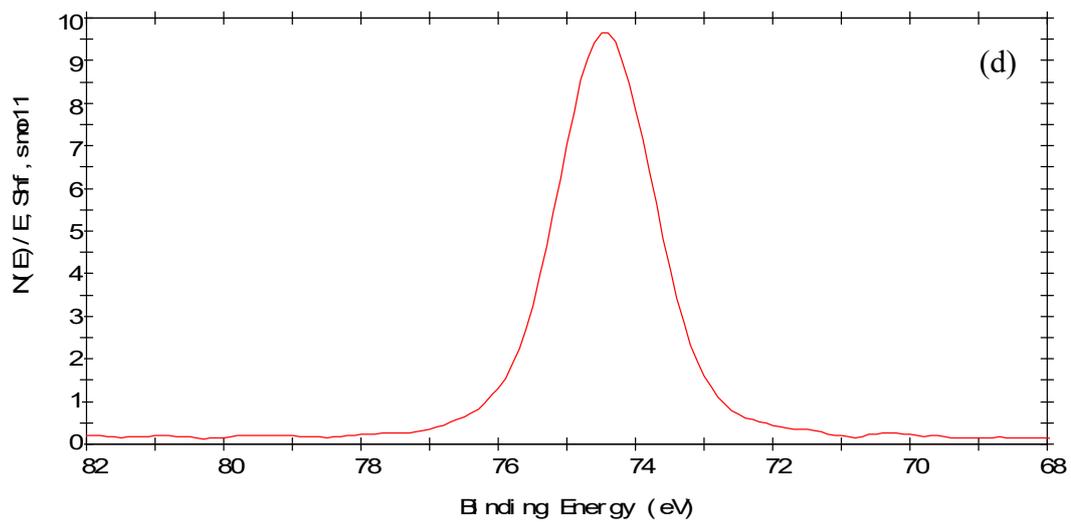
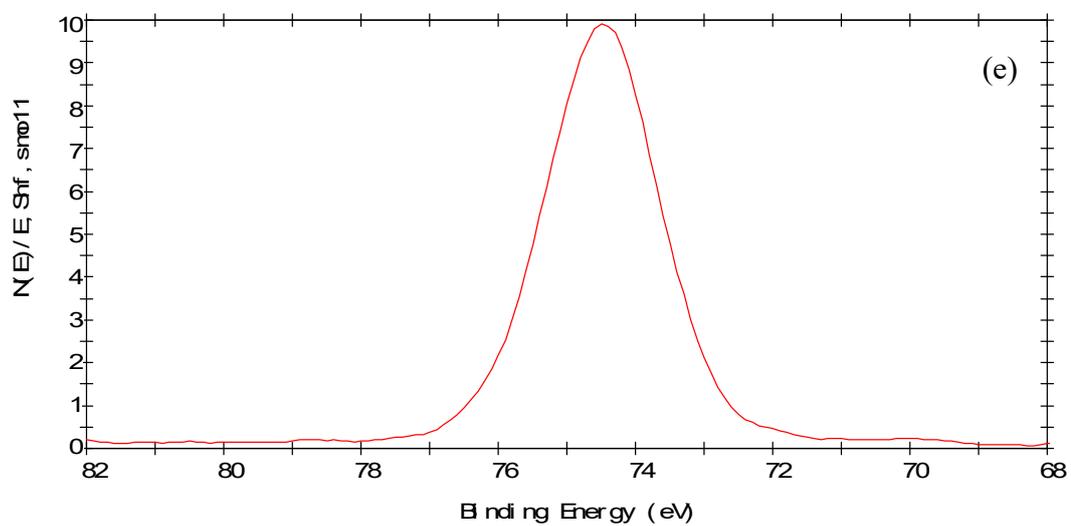
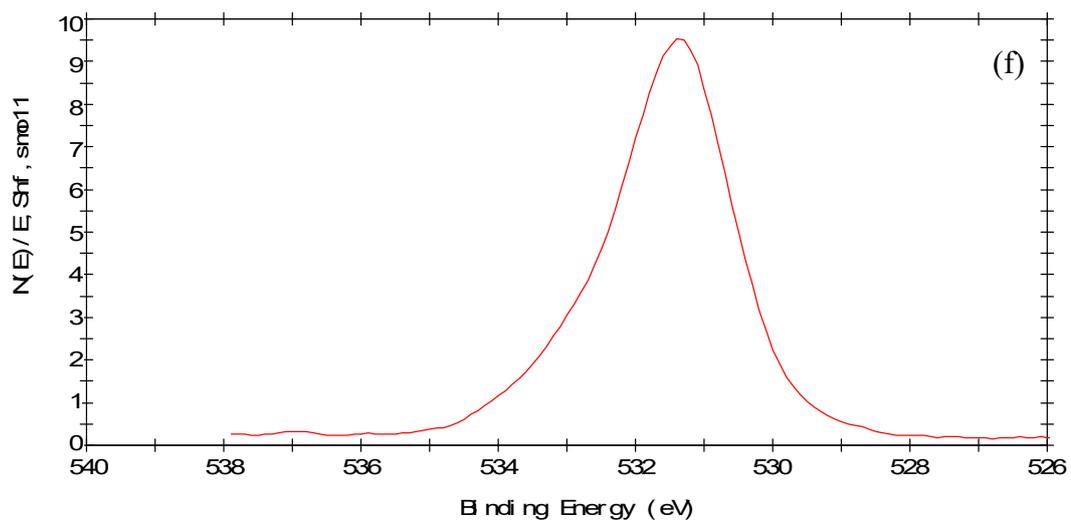

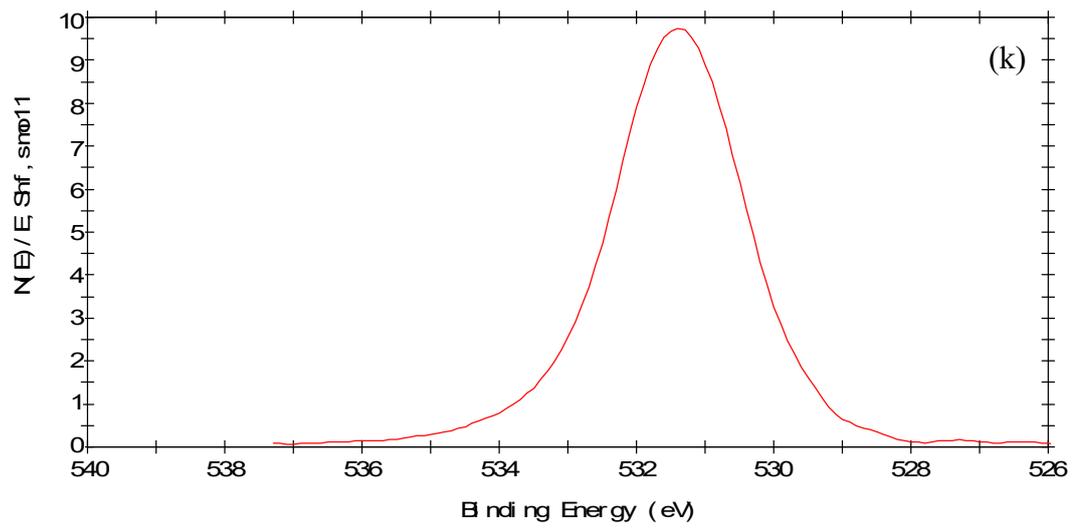

Figure 4

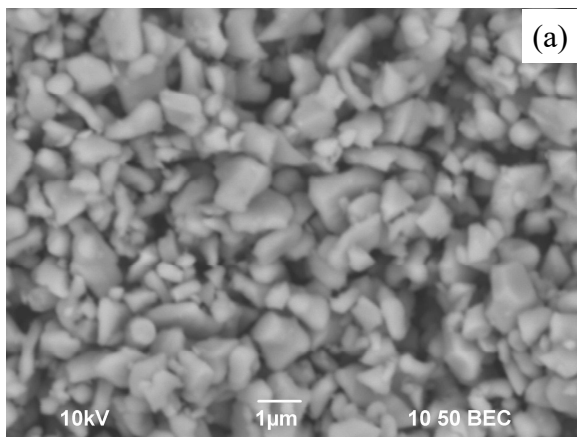
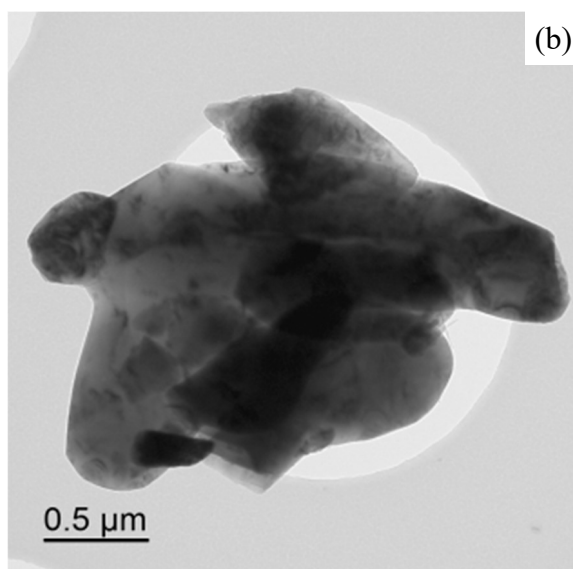
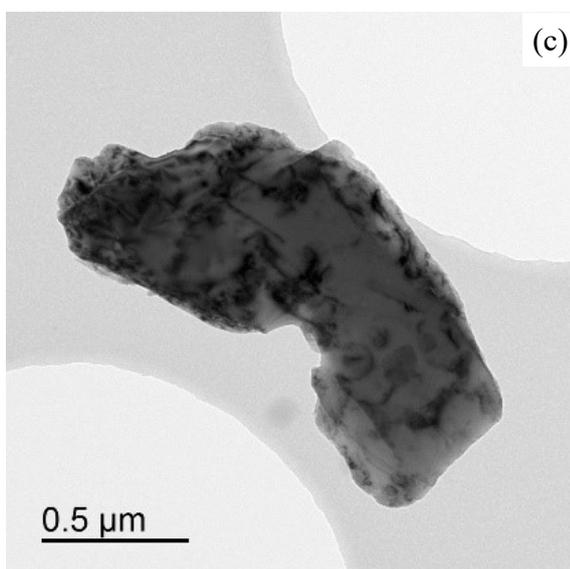

Figure 5

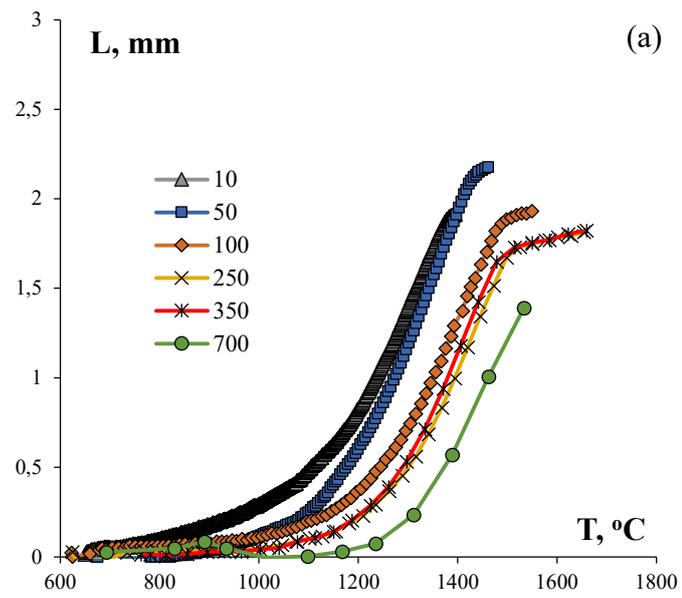
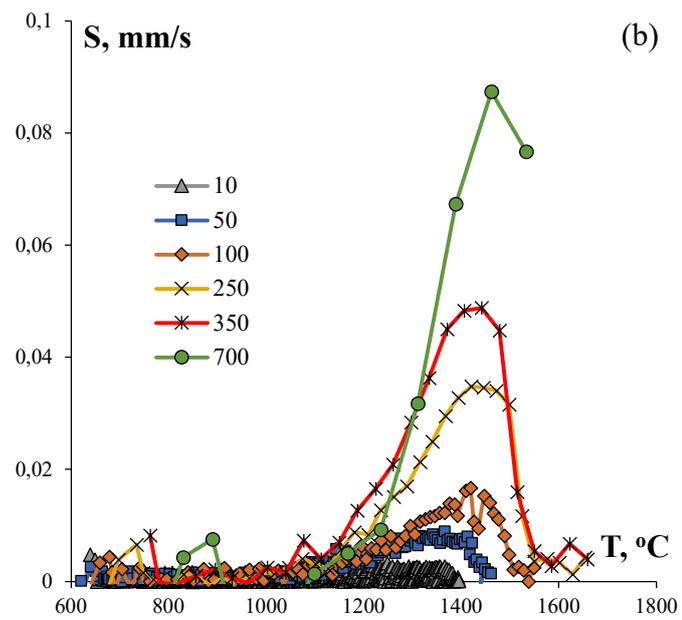

Figure 6

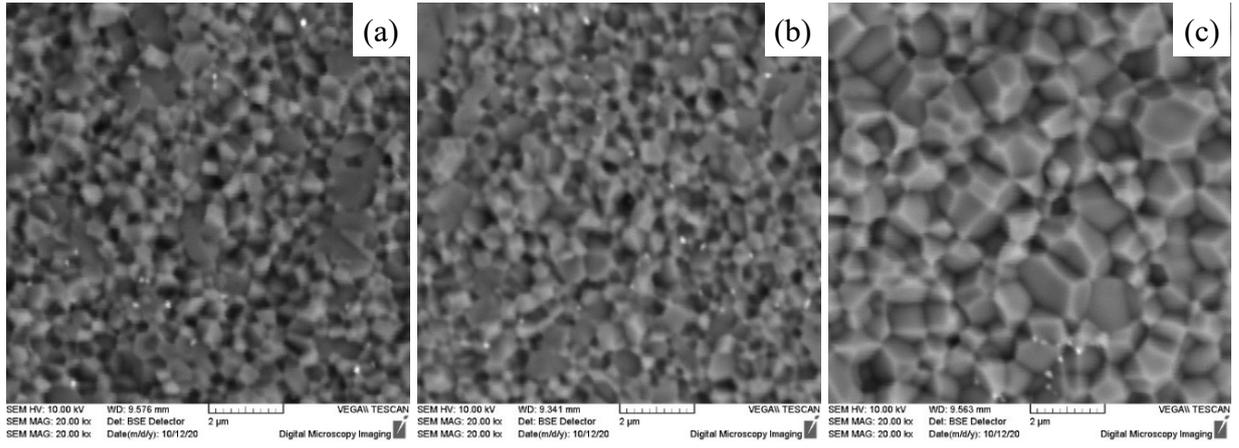

Figure 7

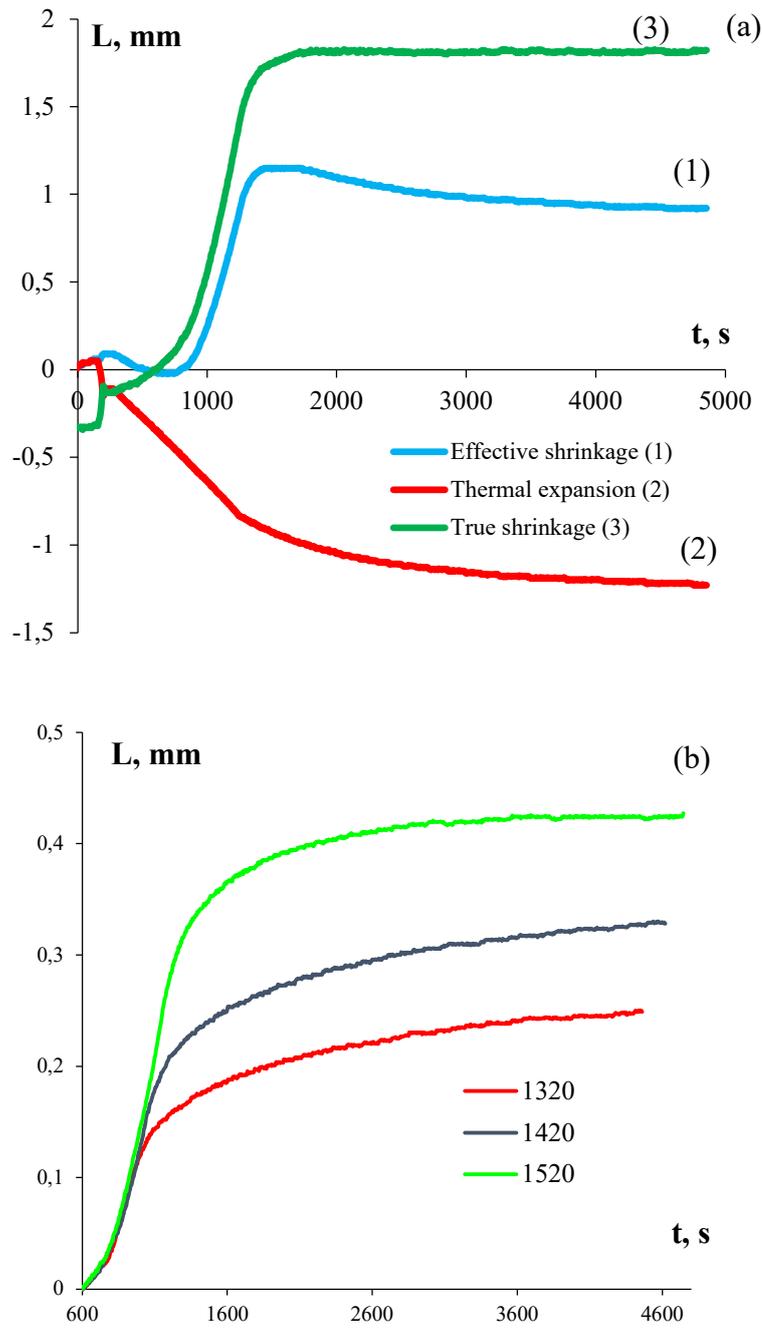

Figure 8

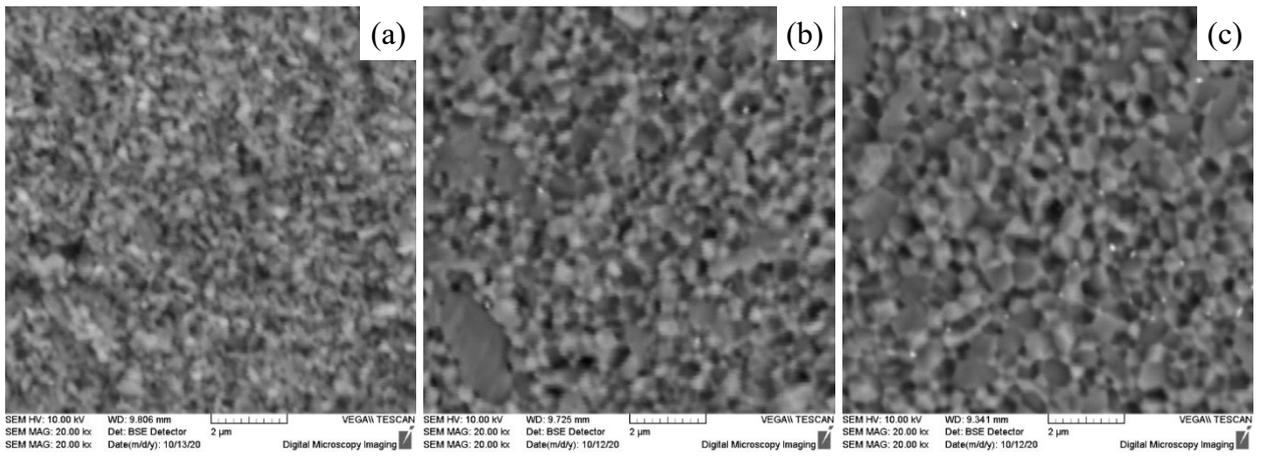

Figure 9

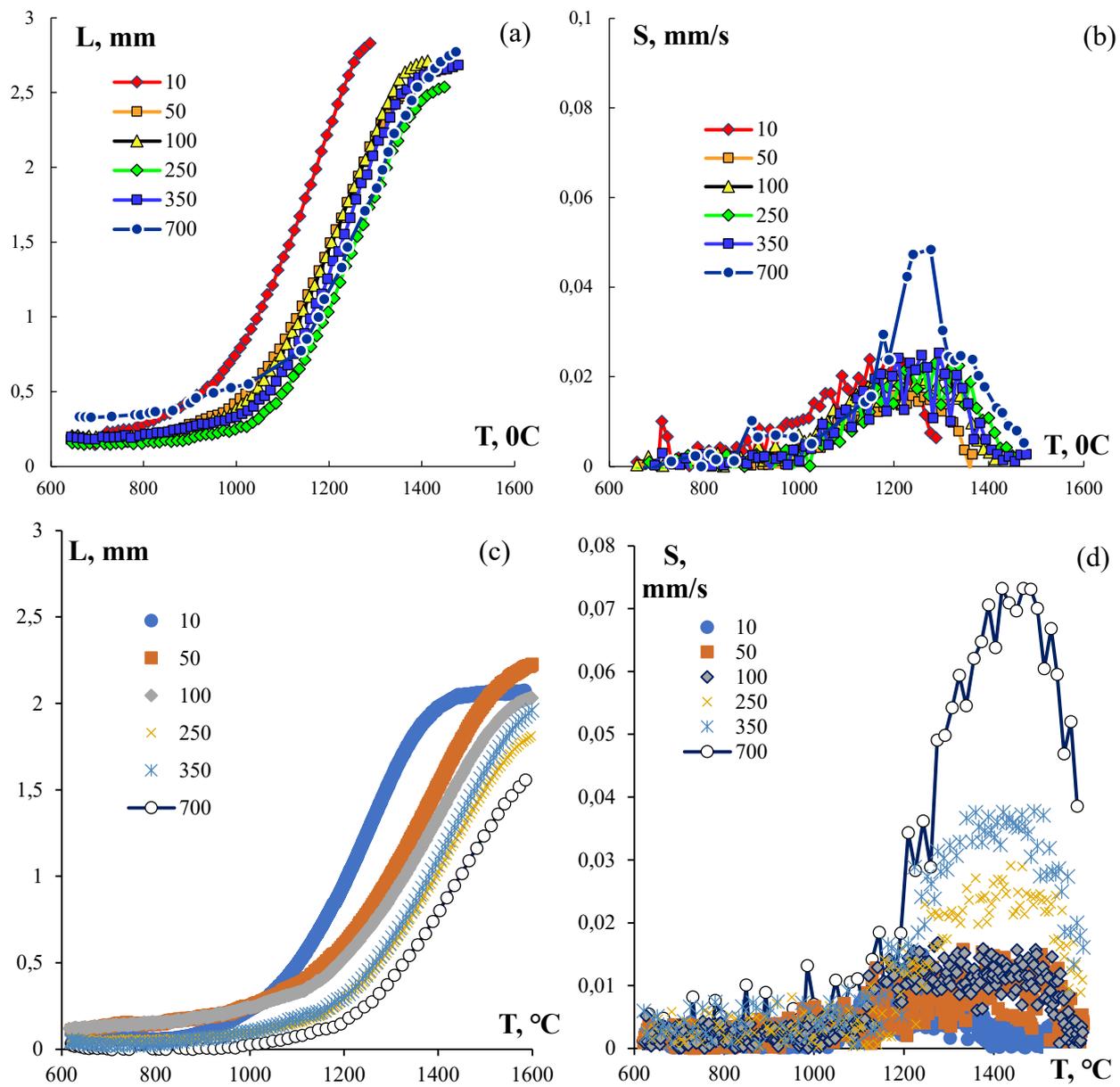

Figure 10

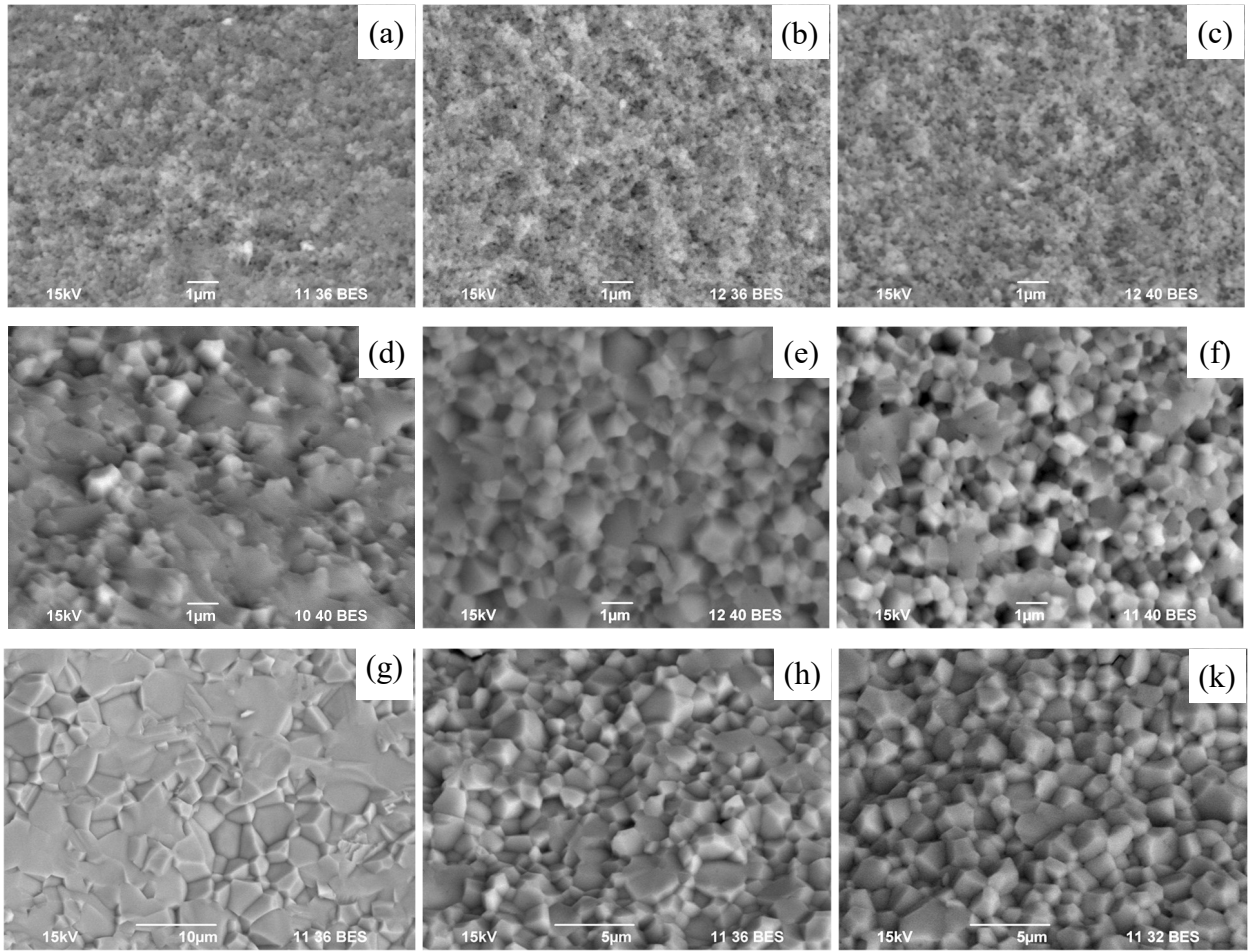

Figure 11

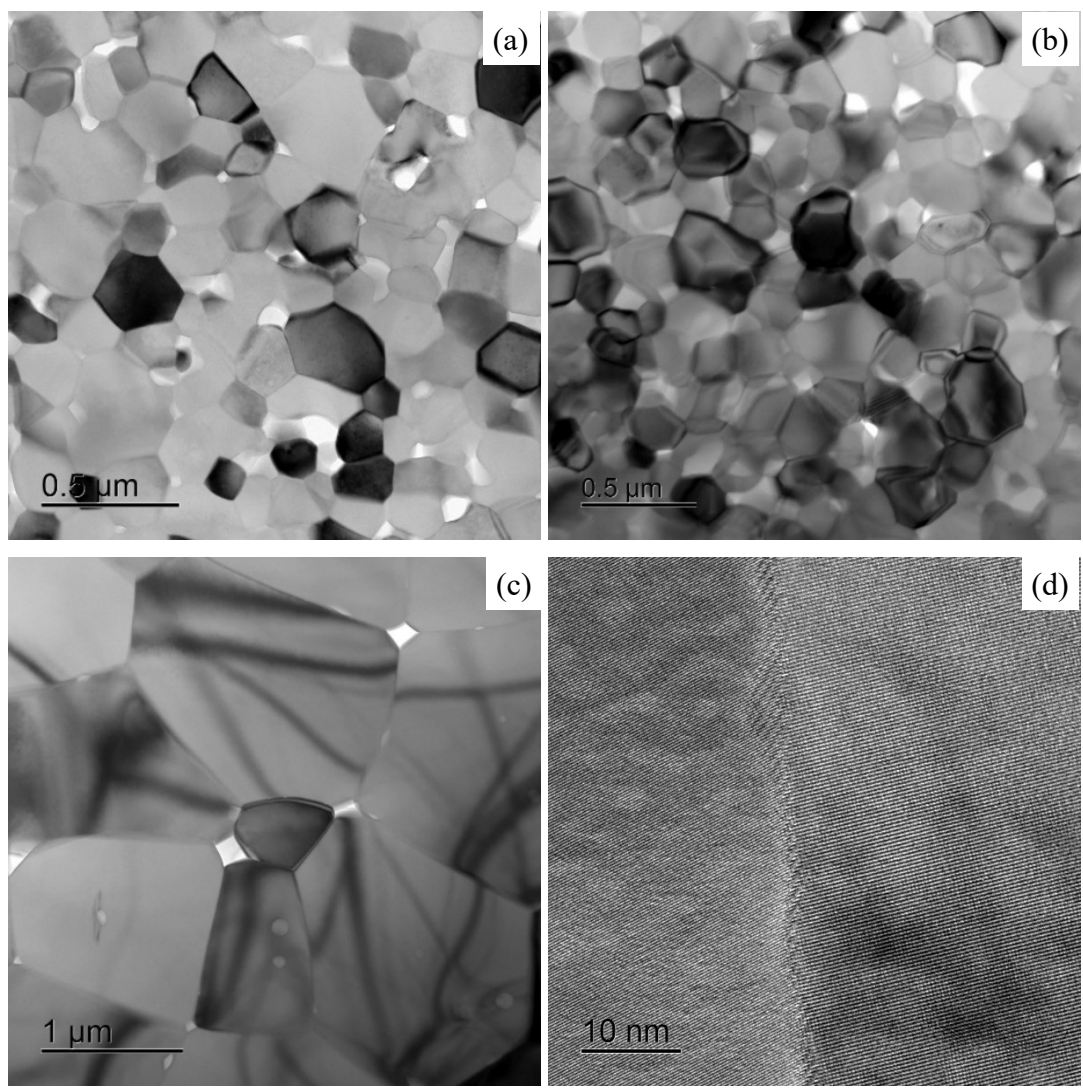

Figure 12

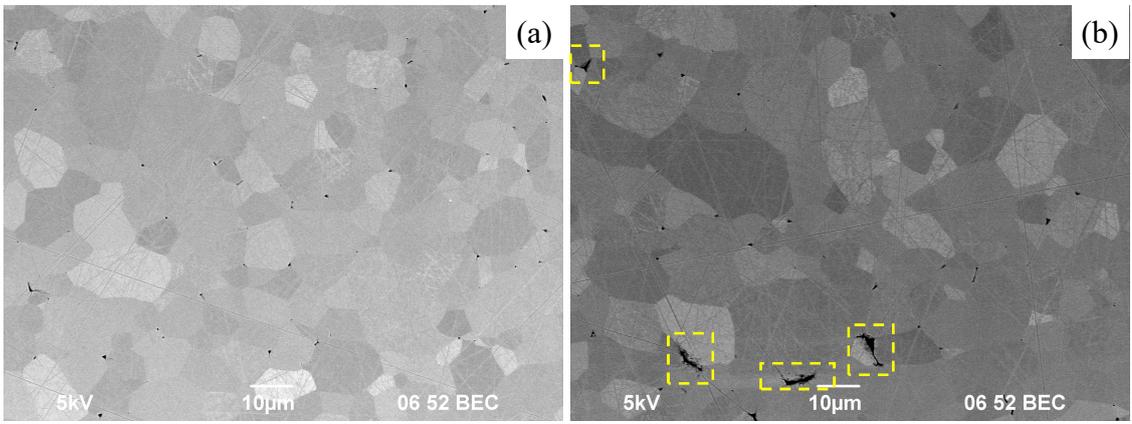

Figure 13

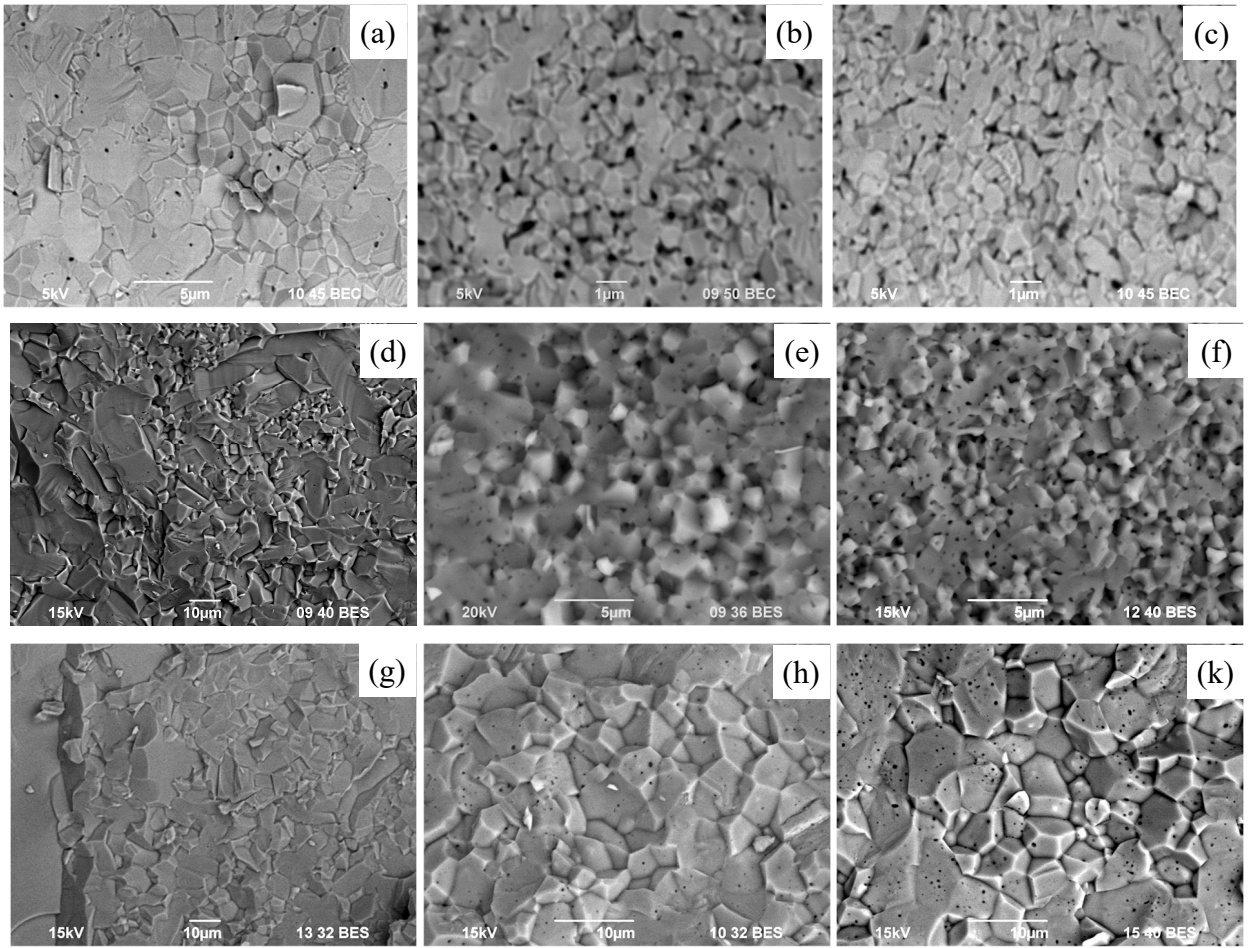

Figure 14

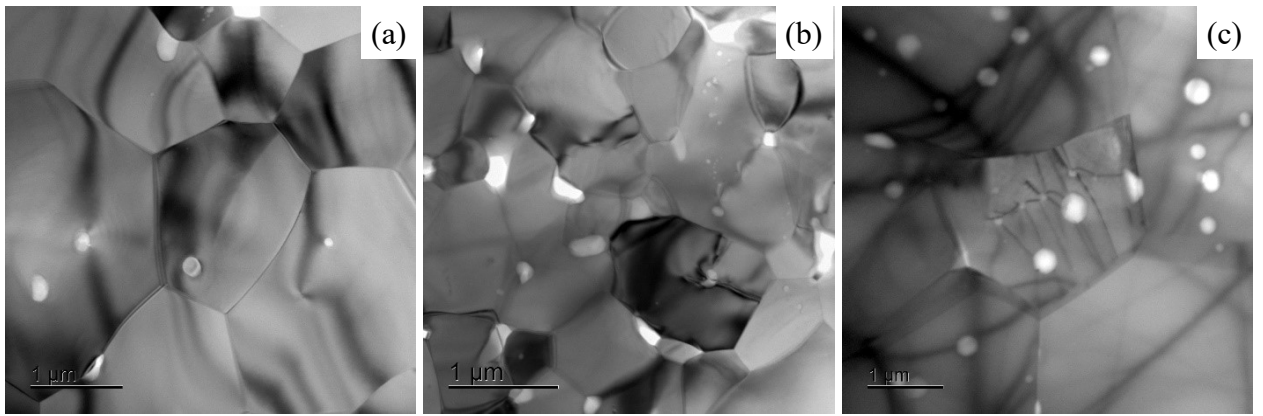

Figure 15

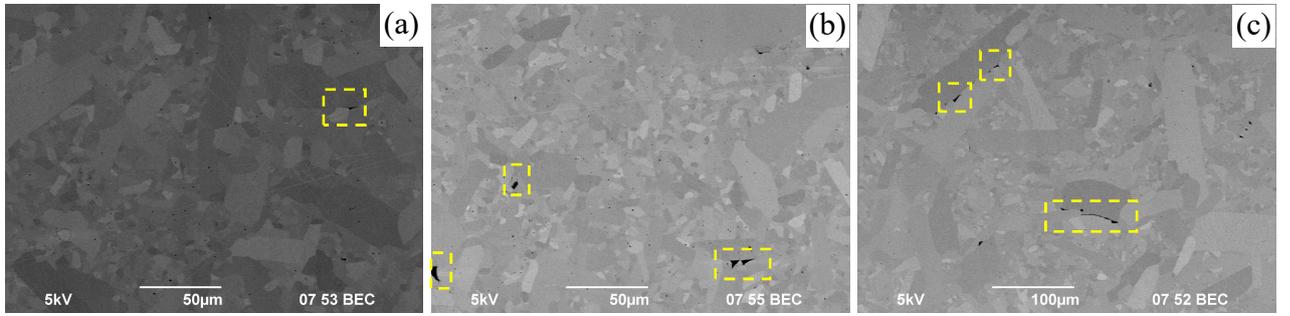

Figure 16

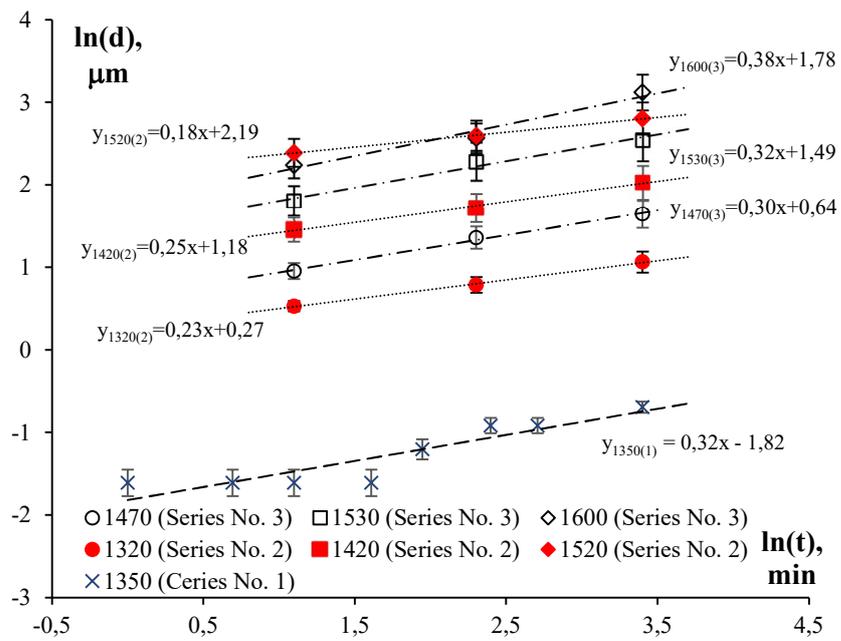

Figure 17

Figure 18

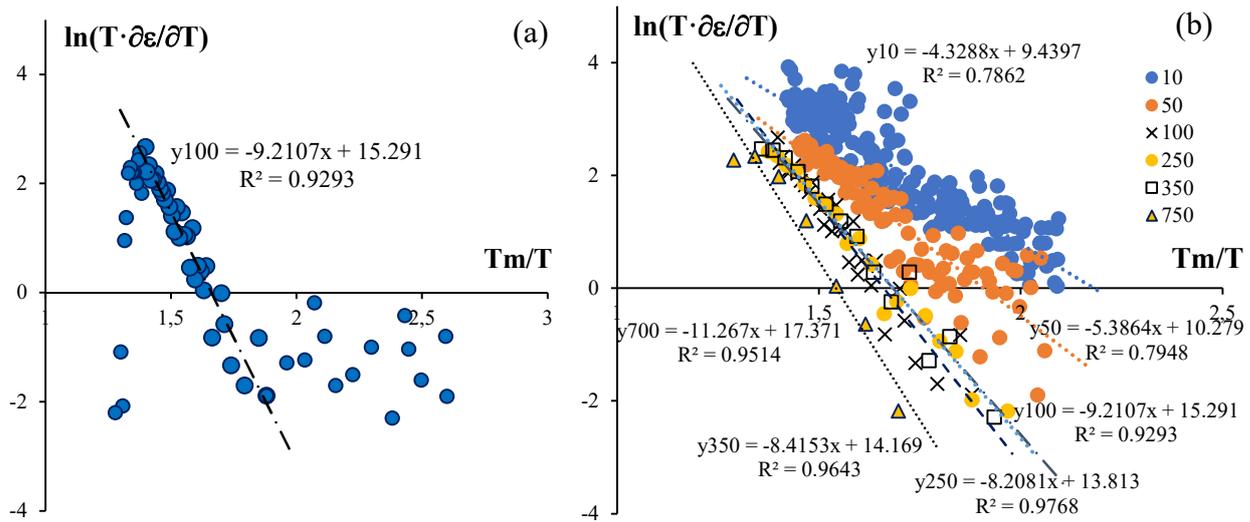

Figure 19

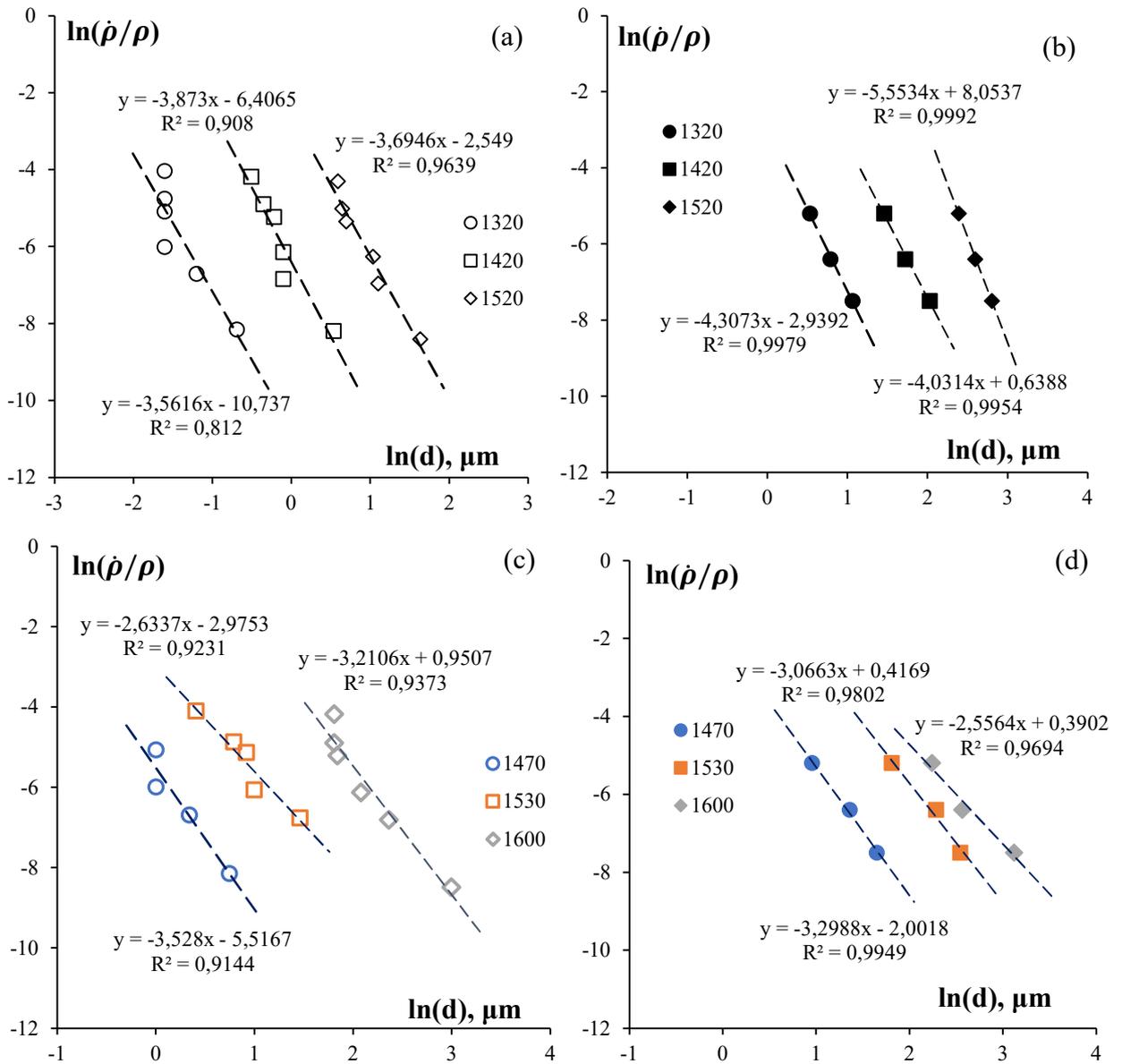

Figure 20